\documentclass[12pt]{article}
\pdfoutput=1

\usepackage[DIV13]{typearea}
\usepackage{indentfirst}
\usepackage{amsmath}
\usepackage{amsfonts}
\usepackage{mathrsfs}
\usepackage{amssymb}
\usepackage{mathtools}
\usepackage{bbm}
\usepackage{epsfig}
\usepackage{graphicx}
\usepackage{subfigure}
\usepackage{slashed}
\usepackage{multicol}
\usepackage[usenames,dvipsnames]{color}
\usepackage{cite}
\RequirePackage[colorlinks=true,urlcolor=blue,anchorcolor=blue,citecolor=blue,filecolor=blue,
               linkcolor=blue,menucolor=blue,linktocpage=true,pdfproducer=medialab]{hyperref}
\usepackage{cancel}
\usepackage{feynmp}
\usepackage{lscape}
\usepackage{multirow}
\usepackage{array}
\usepackage{pdfpages}
\usepackage{fancyhdr}
\usepackage{pifont}

\usepackage[left=.9in, right=.9in]{geometry}

\newcommand{\email}[1]{\href{mailto:#1}{\tt #1}}

\numberwithin{equation}{section}
\newcommand{\LL}{\mathscr{L}}
\newcommand{\cG}{\mathscr{G}}
\def\cB{{\cal B}}

\def\cG{{\cal G}}

\def\cJ{{\cal J}}

\def\cO{{\cal O}}

\def\cT{{\cal T}}

\def\rmU{{\rm U}}

\def\cw{c_{\textrm w}}
\def\sw{s_{\textrm w}}

\def\be{\begin{equation}}
\def\ee{\end{equation}}
\def\beq{\begin{equation}}
\def\eeq{\end{equation}}
\def\bc{\begin{center}}
\def\ec{\end{center}}
\def\bea{\begin{eqnarray}}
\def\eea{\end{eqnarray}}

\def\bry{\begin{array}}
\def\ery{\end{array}}

\def\nn{\nonumber}
\def\nt{\noindent}

\newcommand{\hc}{\text{h.c.}}
\def\fA{\mbox{{\bf M$_{4+5}$}}}
\def\fB{\mbox{{\bf M$_{4+14}$}}}
\def\oA{\mbox{{\bf M$_{1+5}$}}}
\def\oB{\mbox{{\bf M$_{1+14}$}}}

\def\lra#1{\overset{\text{\scriptsize$\leftrightarrow$}}{#1}}

\def\fourpletL{\Psi_{\textbf{4}L}}
\def\fourpletR{\Psi_{\textbf{4}R}}
\def\fourplet{\Psi_{\bf{4}}}
\def\BarfourpletL{\overline{\Psi}_{\textbf{4}L}}

\def\Barfourplet{\overline{\Psi}_{\bf{4}}}

\def\singletL{\Psi_{\textbf{1}L}}
\def\singletR{\Psi_{\textbf{1}R}}
\def\singlet{\Psi_{\bf{1}}}

\def\5qplet{q_L^{\mathbf{5}}}
\def\qBar5plet{\overline{q}^{\bf{5}}_L}
\def\14qplet{q_L^{\bf{14}}}
\def\q14Barplet{\overline{q}^{\bf{14}}_L}

\def\u5plet{u_R^{\bf 5}}
\def\uBar5plet{\overline{u}^{\bf{5}}_R}
\def\usinglet{u_R^{\bf 1}}

\def\d5plet{d_R^{\bf 5}}
\def\dBar5plet{\overline{d}^{\bf{5}}_R}

\def\Xtt{{X_{\hspace{-0.09em}\mbox{\scriptsize2}\hspace{-0.06em}{\raisebox{0.1em}{\tiny\slash}}\hspace{-0.06em}\mbox{\scriptsize3}}}}
\def\Xft{{X_{\hspace{-0.09em}\mbox{\scriptsize5}\hspace{-0.06em}{\raisebox{0.1em}{\tiny\slash}}\hspace{-0.06em}\mbox{\scriptsize3}}}}

\newcommand{\cJmuuichi}{\cJ^{\mu}_{i\chi}}

\newcommand{\cJuichi}{\cJ^\chi_i}
\newcommand{\cJujchi}{\cJ^\chi_j}

\newcommand{\cJmuuaichi}{\cJ^{\mu a}_{i\chi}}
\newcommand{\cJmuuajchi}{\cJ^{\mu a}_{j\chi}}
\newcommand{\cJmuuq}{\cJ^\mu_q}
\newcommand{\cJmuuauchi}{\cJ^{\mu a}_{u\chi}}

\newcommand{\cJmuupsi}{\cJ^\mu_\psi}
\newcommand{\cJmuuqpsi}{\cJ^\mu_{q\psi}}
\newcommand{\cJmuuupsi}{\cJ^\mu_{u\psi}}

\newcommand{\cJmuuqL}{\cJ^{\mu}_{qL}}

\newcommand{\cJmuupsiL}{\cJ^{\mu}_{\psi L}}
\newcommand{\cJmuuqpsiL}{\cJ^{\mu}_{q\psi L}}
\newcommand{\cJmuuupsiL}{\cJ^{\mu}_{u\psi L}}

\newcommand{\cJmuuqR}{\cJ^{\mu}_{qR}}

\newcommand{\cJmuupsiR}{\cJ^{\mu}_{\psi R}}
\newcommand{\cJmuuqpsiR}{\cJ^{\mu}_{q\psi R}}
\newcommand{\cJmuuupsiR}{\cJ^{\mu}_{u\psi R}}

\newcommand{\cJuq}{\cJ_q}

\newcommand{\cJupsi}{\cJ_\psi}
\newcommand{\cJuqpsi}{\cJ_{q\psi}}
\newcommand{\cJuupsi}{\cJ_{u\psi}}

\newcommand{\cJmunuuichi}{\cJ^{\mu\nu}_{i\chi}}

\newcommand{\cJmunuupsi}{\cJ^{\mu\nu}_\psi}
\newcommand{\cJmunuuqpsi}{\cJ^{\mu\nu}_{q\psi}}
\newcommand{\cJmunuuupsi}{\cJ^{\mu\nu}_{u\psi}}

\newcommand{\cJmunuuqu}{\cJ^{\mu\nu}_{qu}}

\newcommand{\rhochi}{\rho_\chi}

\newcommand{\rhochimuu}{\rho^\mu_\chi}
\newcommand{\rhochimud}{\rho_{\mu\chi}}
\newcommand{\rhochimua}{\rho^{a\mu}_\chi}
\newcommand{\rhochinuu}{\rho^\nu_\chi}

\newcommand{\rhoLmu}{\rho^\mu_L}
\newcommand{\rhoRmu}{\rho^\mu_R}

\newcommand{\echimuu}{e^\mu_\chi}
\newcommand{\echimud}{e_{\mu\chi}}

\newcommand{\emunuchimud}{e_{\mu\nu\chi}}

\newcommand{\rhomunuchiu}{\rho^{\mu\nu}_{\chi}}
\newcommand{\rhomunuchid}{\rho_{\mu\nu\chi}}

\newcommand{\BarT}{\overline{T}}

\newcommand{\BarTR}{\overline{T}_R}

\newcommand{\grhochi}{g_{\rho_\chi}}
\newcommand{\mrhochi}{m_{\rho_\chi}}

\newcommand{\aichi}{\alpha^\chi_i}
\newcommand{\ajchi}{\alpha^\chi_j}

\newcommand{\bichi}{\beta^\chi_i}

\newcommand{\blue}[1]{\color{blue} #1 \color{black}}

\newcommand{\eg}{e.g.\,\,}
\newcommand{\ie}{i.e.\,\,}

\newcommand{\hhref}[1]{\href{http://arxiv.org/abs/#1}{arXiv:#1}}

\begin{document}
\begin{titlepage}
\vspace*{-1cm}
\phantom{hep-ph/***} 
\vskip 1cm
\begin{center}
\mathversion{bold}
{\LARGE\bf Top partner-resonance interplay in a Composite Higgs framework}\\
\mathversion{normal}
\vskip .3cm
\end{center}
\vskip 0.5  cm
\begin{center}
{\large Juan Yepes}~$^{a)}$ and
{\large Alfonso Zerwekh}~$^{a)}$
\\
\vskip .7cm
{\footnotesize
$^{a)}$~
\emph{Department of Physics and Centro Cient\'{i}fico-Tecnol\'{o}gico de Valpara\'{i}so\\
Universidad T\'{e}cnica Federico Santa Mar\'{i}a, Valpara\'{i}so, Chile}\\
\vskip .1cm
\vskip .3cm
\begin{minipage}[l]{.9\textwidth}
\begin{center} 
\textit{E-mail:} 
\email{juan.yepes@usm.cl},
\email{alfonso.zerwekh@usm.cl}
\end{center}
\end{minipage}
}
\end{center}
\vskip 0.5cm
\begin{abstract}
\nt Guided us by the scenario of weak scale naturalness and the possible existence of exotic resonances, we have explored in a $SO(5)$ Composite Higgs set-up the interplay among three matter sectors: elementary, top partners and vector resonances. We parametrise it through explicit interactions of spin-1 $SO(4)$-resonances, coupled to the $SO(5)$-invariant fermionic currents and tensors presented thoroughly by the first time in this work. Such invariants are built upon the Standard Model fermion sector as well as top partners sourced by the unbroken $SO(4)$. The  mass scales entailed by the top partner and vector resonance sectors will control the low energy effects emerging from our interplaying model. Its phenomenological impact and parameter spaces have been considered via flavour-dijet processes and electric dipole moments bounds.
Finally, the strength of the Nambu-Goldstone symmetry breaking and the extra couplings implied by the top partner mass scales are  measured in accordance with expected estimations.
\end{abstract}
\emph{Keywords:} Composite particle models; Nambu-Goldstone; Top quarks; intermediate bosons;  electric dipole moment.\\

\noindent PACS Nos.: 12.60.Rc; 14.80.Va; 14.65.Ha; 13.38.-b;  31.30.jn.

\end{titlepage}
\setcounter{footnote}{0}

\tableofcontents

%
%

\newpage

\section{Introduction}

\nt The Standard Model (SM) has been successfully and firmly established as a consistent framework of electroweak symmetry breaking (EWSB) after the LHC's Higgs discovery~\cite{Aad:2012tfa,Chatrchyan:2012xdj}, opening thus new windows for exploring BSM scenarios aimed at healing the long-standing Hierarchy Problem. One of those alternatives postulates a Higgs particle emerging from a strong composite sector at the TeV or slightly higher scale, being thus shielded from the UV physics by its composite nature and distinguished from the other composite resonances due to the Nambu-Goldstone (NG) symmetry. These features have shaped the so-called Composite Higgs models (CHMs)~\cite{Kaplan:1983fs,Kaplan:1983sm,Georgi:1984ef,Banks:1984gj,Georgi:1984af,Dugan:1984hq,Contino:2003ve,Agashe:2004rs,Contino:2010rs}, where the measured value of the Higgs mass demands the existence of lighter exotic resonances, concretely, colored composite fermionic resonances with a mass below 2~TeV~\cite{Contino:2006qr,Matsedonskyi:2012ym,Marzocca:2012zn,Pomarol:2012qf,Redi:2012ha,Panico:2012uw}.
As a bonus, these CHMs allow large number of New Physics (NP) signatures concerning direct production at the LHC~\cite{Contino:2008hi,Mrazek:2009yu}, as well as indirect NP probes, such as flavour and  electroweak precision tests (EWPT) observables~\cite{Matsedonskyi:2014iha}. Complementary, these models have been consistently armed with exotic spin-0 and spin-1 resonances at the TeV scale, whose impact on the pseudo NG bosons (PNGBs) scattering, and then on the high-energy vector boson scattering, have been thoroughly studied~\cite{Contino:2011np}.

Motivated by the challenging scenario of weak scale naturalness, the aim at this work is to explore the low energy implications from the interplay among three matter sectors: elementary, composite partners and spin-1 resonances in a $SO(5)$ CHM. We have parametrised such interactions through couplings of the vector resonance $\rho$, here assumed to be spin-1 triplets of $SU(2)_L\times SU(2)_R$, to a set of $SO(5)$-invariant fermionic currents and tensors presented in this analysis. Such invariants cover all the structures built upon the SM elementary sector together with the top partners $\Psi$ permitted by the unbroken $SO(4)$. All this matter content will frame four models, coupled each of them to the vector resonance $\rho$, and with their low energy effects subsequently scanned along some regimes of the implied NG boson scale.

The top partners and vector resonances will play a role below the cut-off of our theory at the mass scales $M_\Psi$ and $m_\rho$ correspondingly. Low energy effects sourced by our interplaying picture will be tackled by assuming the hierarchy $M_\Psi < m_{\rho}$. Via the associated equations of motion (EOM) for $\rho$ and $\Psi$, we will be led to obtain 4-fermion operators whose phenomenological impact and parameter spaces will be considered via flavour-dijet processes and EDM bounds. Parametric spaces implied by the coefficients weighting the current and tensors were explored and turned out to be of sizeable order of magnitude. Finally, we will be able to measure the strength of the NG symmetry breaking and the extra couplings provided by the top partner mass scales, accordingly with theoretical expectations.

Top quark physics at CHMs have been extensively studied ~\cite{Marzocca:2012zn,DeSimone:2012fs,ewpt}, with general flavour physics analyses~\cite{Barbieri:2012tu,Redi:2012uj,KerenZur:2012fr}, considered in the shed light of top partner sectors~\cite{Matsedonskyi:2014iha}, whilst spin-0 and spin-1 resonances have been considered in CHMs~\cite{Contino:2011np} with updated analysis~\cite{CarcamoHernandez:2017pei,Hernandez:2015xka}. Our discussion will be based on the previous studies~\cite{Matsedonskyi:2014iha,DeSimone:2012fs}, extended up to a simple approach accounting for effective top partners-vector resonances interplay and its low energy consequences.

This work is split into: CCWZ construction for the CHM and  introduction of the composite fermions in Sections~\ref{Elementary-composite sectors}-\ref{Top-partners-fields}. Vector resonance sector and its couplings to the elementary-composite sector in Sections~\ref{Vector-resonance-Lagrangian}-\ref{Interplay}. Low energy consequences, some flavour and dijet bounds in Section~\ref{Low-energy}. We briefly comment on the production channels for the top partners and vector resonances in Section~\ref{Production}, finally summarising all the work in Section~\ref{Summary}.

\section{Elementary-composite sectors}
\label{Elementary-composite sectors}

\nt The underlying CHM is made out of two sectors:

\begin{itemize}

\item \emph{The composite sector}, containing the composite Higgs boson and other composite resonances. The Higgs is introduced via the general CCWZ formalism~\cite{ccwz} as a PNGB of the minimal global symmetry $\cG=SO(5)$~\cite{Agashe:2004rs} and spontaneously broken to $SO(4)$ by the strong sector at the scale $f$. Four massless PNGBs are generated, yielding thus an $SU(2)_L$ Higgs doublet\footnote{Hence the Higgs is exactly massless unless the strong sector is coupled to some source of an explicit $\cal G$-breaking.}. An additional $U(1)_X$ factor is introduced in order to reproduce the proper SM hypercharge $Y=T_R^3+X$, then $\cG=SO(5)\times U(1)_X$. The PNGBs enter through the  $5\times 5$ Goldstone matrix
\be
\rmU=\exp\left[i \frac{\sqrt{2}}{f}\,\Pi^i\,T^i\right]=\,
\left(\begin{array}{ccccc}
& & \vspace{-3mm}& & \\
 & \mathbb{I}_{3} & & &  \\
  & & \vspace{-3mm}& &  \\ 
   & & & \cos \frac{h + \langle h \rangle}{f} & \sin \frac{h + \langle h \rangle}{f} \\
    & & & -\sin \frac{h + \langle h \rangle}{f} & \cos \frac{h + \langle h \rangle}{f} 
\end{array}\right)\,,
\label{GB-matrix}
\ee

\nt where $T^i$ are the coset $SO(5)/SO(4)$-generators, whilst $\Pi^i$ and $f$ are the PNGB fields and the decay constant respectively. Henceforth $T^i$ will stand for the coset generators, while $T^a$  for the unbroken ones, all them defined in~Appendix~\ref{CCWZ}.

\item The \emph{elementary sector}, containing copies of all the SM field sector except for the Higgs transforming under the SM gauge symmetry group ${\cal G}_{\text{SM}} \subset {\cal G}$. This sector is not $\cal G$ invariant, therefore the one-loop effective potential triggered by the elementary-composite interactions allows the Higgs to pick a mass, fixing thus its vacuum expectation value (VEV) in a ${\cal G}_{\text{SM}}$-breaking direction.
\end{itemize}

\nt The SM  symmetry ${\cal G}_{\text{SM}}=SU(2)_L \times U(1)_Y$ is embedded into the unbroken $SO(4)\times U(1)_X$ and its breaking will be triggered via a non-zero Higgs VEV $\langle h \rangle \simeq v=246$~GeV, measuring together with the $SO(5)$ breaking scale $f$, the degree of tuning of the scalar potential through the ratio~\cite{Agashe:2004rs}
\be
\xi=\frac{v^2}{f^2}.
\label{Xi}
\ee

\nt Generically, the value of $f$ must be large to suppress NP effects, but not too far from $v$ to maintain a tolerable tuning. Since $\xi$ controls low energies SM departures, then it cannot be too large. Electroweak precision tests suggest $\xi=\{0.1,\,0.2,\,0.25\}$ which corresponds to $f\approx\{800,\,550,\,490\}$GeV and these will be the values to be tested in this work.  More stringent constraints on $\xi$ have been reported previously, following the current 95\% combined limit from direct production of either the charged $\rho^\pm$ or the neutral $\rho^0$ at the LHC~\cite{Contino:2013gna}. Those limits allow $\xi\sim 0.02$, or even smaller, for a vector resonance mass $m_\rho\sim 2$ TeV. Such small values might be directly tested through single Higgs production at the LHC, reaching larger precision via double Higgs processes at CLIC\footnote{In~\cite{Contino:2013gna} have been reported precision values for $\xi$ according to double Higgs at CLIC, and compared with respect to single and double Higgs production at LHC, single Higgs at ILC  and single Higgs at CLIC as well. Such anaysis infers corresponding maximal bounds on compositeness scale, some of them compatible with the one obtained from constraints on non-standar Higgs couplings (see~\cite{Ciuchini:2013pca}). In a specific model like the Minimal Composite Higgs model the couplings $a$, $b$ and $d_3$ depend on the single parameter $\xi = (v/f)^2$. Through the study of $e^+e^-\to hh\nu\bar\nu$, a machine like CLIC with $\sqrt{s}=3\,$TeV and $L=1\,\text{ab}^{-1}$ can reach a sensitivity as small as $0.02$ on $\xi$. These sensitivities can be translated into an indirect reach  on the cutoff scale $\Lambda \equiv 4\pi f$, that is the mass scale of the resonances for  the case where the underlying dynamics is maximally strong. The values of $\Lambda$ provided in~\cite{Contino:2013gna}, are in agreement with non-standard Higgs couplings measurements at~\cite{Ciuchini:2013pca} and corresponds to $\xi=0.02-0.05$.}, and should be compared with indirect bounds from EW precision data. In fact, by including only tree-level contributions to $\Delta \hat S= m_W^2/m_\rho^2$ from the $\rho$ exchange~\cite{Contino:2011np} and the 1-loop IR effect from the modified Higgs  couplings, it is possible to exclude at $95\%$ the region $\xi\gtrsim 0.03$, with $\xi$ tending to $\sim 0.02$ in the infinite $\rho$-mass case. Having no other contributions to the oblique parameters, masses $m_\rho \lesssim 5\,$TeV are already excluded even for very small $\xi$ (see~\cite{Contino:2013gna} for more details).  Nonetheless, slight modifications to the EW parameter $\hat T$  shift the 95\% exclusion boundary in such a manner that masses as low as $m_\rho \sim 2$~TeV are still viable. On the other hand,  the  parameter $\hat S$ is sensitive both to the cutoff and to the composite resonances scale, thus the inclusion of light fermionic resonances may relax those stringent constraints,  reaching for instance $\xi\sim 0.3$ or even bigger for $m_\rho\sim 3$~TeV in minimal models with fermionic fourplet resonances~\cite{Matsedonskyi:2014iha}. More exotic scenarios, like the nineplet case, would lead to stringent values as $\xi\sim 0.02$ for $m_\rho\sim 3$~TeV in agreement with $99\%$CL bounds from the $\hat S$ parameter~\cite{Matsedonskyi:2014iha}. For the present work we will assume $\xi=\{0.1,\,0.2,\,0.25\}$, as they are compatible with the latter EWPT bounds, and with the vector resonance direct production bounds at LHC, as well as the expected single Higgs production at the LHC, and the double Higgs production at CLIC. In addtion, those values are inside the domain of validity of the scenario, $g_\rho<4\pi$ and they will be assumed henceforth.

Both of the elementary-composite sectors and their interactions will be generically described in this analysis via
\be
\LL = \LL_{\text{elem}}\,\,+\,\,\,\LL_{\text{comp}}\,\,+\,\,\,\LL_{\text{mix}}.
\label{Lagrangian}
\ee

\nt $\LL_{\text{mix}}$ is responsible for the breaking of the Goldstone symmetry through the partial compositeness mechanism, the which postulates an UV Lagrangian above the $\cal G$ symmetry breaking scale containing linear couplings between elementary fermions $q$ and strong sector operators dictated by
\be
{\cal L}_{\text{mix}}^{\text{UV}} = \sum_q y\,\bar q\,\cO_q .
\label{UV-mix}
\ee

\nt The operators $\cO_q$ transform in one of the $SO(5)$ representations, shaping as well the corresponding embeddings of the elementary fields. We will consider two choices:

\begin{itemize}

\item Fundamental ${\bf 5}$ representation, with both chiralities of the fermion $q$ having elementary representatives coupled to the strong sector through the ${\bf 5}$-plets
\be
\hspace*{0.5cm}
\5qplet={1\over \sqrt{2}}\left(\begin{matrix}
i d_L\\
 d_L\\
i u_L\\
- u_L\\
0
\end{matrix}\right),
\quad
\u5plet = \left(
\begin{array}{c}
0\\
0\\
0\\
0\\
u_R
\end{array}
\right),
\label{emb}
\ee

\item Fundamental ${\bf 14}$ representation, with the right-handed $q$ quark as a totally composite state arising itself from the operator ${\cal Q}_q$ at low energies, coupled to $q_L$ by means of the mixing~\eqref{UV-mix}. The fields coupled to the strong sector are
\be
\hspace*{0.5cm}
\14qplet={1\over \sqrt{2}}\left(\begin{matrix}
0 & 0 & 0 & 0 & i d_L\\
0 & 0 & 0 & 0 & d_L\\
0 & 0 & 0 & 0 & i u_L\\
0 & 0 & 0 & 0 & -u_L\\
i d_L & d_L & i u_L & -u_L &0\\
\end{matrix}\right),
\quad
\usinglet\,.
\label{emb}
\ee

\end{itemize}

\nt In both cases the representations $q_L$ and $u_R$ have the same $X$-charge $2/3$, allowing to reproduce the correct electric charge of the top. The doublet $q^T_L=(u_L,d_L)$ has an isospin $T_R^3=-1/2$, providing thus a protection from large deformations of the $b_L$-couplings~\cite{Agashe:2006at,Mrazek:2011iu}. The top partners hence have the same $X$-charge equal to 2/3 as the  composite operator ${\cal O}_q$ and are introduced in the next.
 
The couplings in~\eqref{UV-mix}, though explicitly break the $\cG=SO(5)\times U(1)_X$ symmetry, must of course  respect the SM group. Restoring $\cG$ at the elementary-composite interactions of~\eqref{UV-mix} demands suitable transformation properties for the embeddings. Under $g\in{\textrm{SO(5)}}$ 
we have
\be
\left(\5qplet\right)_i\;\rightarrow g_i^{\,j}\left(\5qplet\right)_j\,,\qquad\left(\14qplet\right)_{i\,j}\;
\rightarrow g_i^{\,l} g_j^{\,k}\left(\14qplet\right)_{l\,k}\,,
\label{transemb}
\ee
while the ${\textrm{U(1)}_X}$-charge is equal to $2/3$ in both cases. Such symmetry transformation is accounted for building up the interplaying currents later on that will lead to the emergence of EFT 4-fermion operators in our framework.

\section{Top partners fields}
\label{Top-partners-fields}

\nt A fourplet $\fourplet$ and a singlet $\singlet$ representations are naturally sourced by the decomposition rule ${\bf 5}={\bf 4}+{\bf 1}$ under the unbroken $SO(4)$ group and can be encoded through 
\be
\fourplet={1\over \sqrt{2}}\left(\begin{matrix}
i\cB-i\Xft\\
\cB+\Xft\\
i\cT+i\Xtt\\
-\cT+\Xtt
\end{matrix}\right),\qquad\qquad \singlet=\widetilde{\cT}\,.
\label{fourplet-singlet}
\ee

\nt The fourplet $\fourplet$ is decomposable into two doublets $(\cT,\cB)$ and $(\Xft,\Xtt)$ of hypercharge $1/6$ and $7/6$ 
respectively. The former has the same quantum numbers as the doublet $(u_L,d_L)$, whilst the latter contains a state of exotic charge $5/3$ plus another top-like quark $\Xtt$. On the other hand, the singlet representation $\singlet$ entails only one exotic top-like state, denoted in here as $\widetilde{\cT}$. 

Following a slightly modified notation as in~\cite{DeSimone:2012fs}, the cases of a strong sector operators $\cO_q$ as $\bf{5}$ or $\bf{14}$, in the presence of either a fourplet or singlet partner, will shape the models 
\be
\bf M=\bf M_{\Psi +q}=\{\fA,\,\fB,\,\oA,\,\oB\}\,.
\label{Models}
\ee

\nt Note that by decomposing $\cO_q$ in~\eqref{UV-mix} under the unbroken $SO(4)$ we obtain, respectively, ${\bf 5}={\bf 4}+{\bf 1}$ and ${\bf 14}={\bf 4}+{\bf 1}+{\bf 9}$, being possible to find thus a fourplet and/or a singlet in the low-energy spectrum. The nineplet case $\bf{9}$ is beyond the scopes of the present work and it is left for a future analysis~\cite{YZ}.

In the next we briefly introduced the spin-1 vector resonances that will be coupled a posteriori to the fermion field content described so far.

\section{Vector resonance Lagrangian}
\label{Vector-resonance-Lagrangian}

\nt Below the cut-off of the theory at $\Lambda=4\pi f$ we assume the existence of spin-1 resonances parametrized by a mass $m_\rho\simeq g_\rho f$ and a coupling $1< g_\rho < 4\pi$, who controls both the interactions among the resonances and the resonance-pion interactions. 

We are concerned here only to the case of resonances transforming under $SO(4)$. According to the rule $\mathbf{4}\times \mathbf{4} = \mathbf{1}+ \mathbf{6} +\mathbf{9}$ the resonance can therefore be encoded by one of the $SU(2)_L\times SU(2)_R$-representations $(\mathbf{1},\mathbf{1}) + (\mathbf{3},\mathbf{1}) + (\mathbf{1},\mathbf{3}) + (\mathbf{3},\mathbf{3})$. For the work undertaken in here, only the spin-1 resonances $\rhoLmu=(\mathbf{3},\mathbf{1})$ and $\rhoRmu =(\mathbf{1},\mathbf{3})$ will be considered. Their description as triplet representations of $SU(2)_L\times SU(2)_R$ follows the well known vector formalism~\cite{Ecker:1989yg}, where the fields transform non-linearly under a transformation $\Pi\in SO(5)$ as
\be
\rhochimuu = \rhochimua\,T^a_\chi,\qquad \rhochimuu \rightarrow {\bf h}\,\rhochimuu\,{\bf h}^{\dagger}\,+\,\frac{i}{\grhochi}\,\left({\bf h}\,\partial^\mu {\bf h}^{\dagger}\right)_\chi\,,
\label{Resonance-transformation}
\ee

\nt with the notation $\chi=L,\,R$, and ${\bf h}=h(\Pi,h^{\hat a})$. The $SO(4) \simeq SU(2)_L \times SU(2)_R$ unbroken generators $T^a_\chi$ are defined in~\ref{CCWZ}. It is customary to write the effective Lagrangian for the spin-1 resonance as
\be
\LL_{\rho_\chi}=-\frac{1}{4\,\grhochi^2}\,\rhomunuchiu\rhomunuchid\,\,\,\,+\,\,\,\,\frac{\mrhochi^2}{2\,\grhochi^2}\left(\rhochimuu-\echimuu\right)^2
\label{rho-Lagrangian}
\ee

\nt where an internal sum over the $SO(4)$ unbroken generators indices is assumed through the strength field and vector products in~\eqref{rho-Lagrangian}, with the CCWZ symbol $\echimuu$ defined in~\ref{CCWZ} and the strength field written as usual 
\be
\rhomunuchiu = \partial^\mu \rhochinuu\,\,-\,\,\partial^\nu \rhochimuu\,\,+\,\,i\left[\rhochimuu , \rhochinuu\right].
\ee 

\nt Higher-order operators $Q_i$ (in field and derivative) may expand the LO Lagrangians in~\eqref{rho-Lagrangian} and are disregarded for the present work (see~\cite{Contino:2011np} more details). Finally, it is worth to remark the similarity between the Lagrangian $\LL_{\rho_\chi}$ in~\eqref{rho-Lagrangian} and the one originally considered in $SO(4)/SO(3)$ models of Hidden Local Symmetry (HLS)~\cite{Bando:1984ej,Casalbuoni:1985kq,Casalbuoni:1986vq,Georgi:1989xy}.

The interplaying interactions among the top partners, elementary fermion fields and the aforementioned vector resonances is considered via an effective approach in the next section.

\section{Fermion--vector interplaying Lagrangian}
\label{Interplay}

\nt So far the fermion sector accounted by the top partners and the composite operators have been separately considered from the previous vector-like formalism. Covering all the possible couplings among them requires the following set of $SO(5)$-invariant interplaying interactions, sourced by the simultaneous presence of both of the sectors, and succinctly described by the whole fermion-vector interplaying Lagrangian
\be
\LL_{\text{int}}=\LL_{\bf M}\,\,+\,\,\sum_{\chi=L,R}\left(\LL_{\rho_\chi}\,\,+\,\,\LL_{\bf M\,+\,\rho_\chi}\,\,+\,\,\LL^\text{mag}_{\bf M\,+\,\rho_\chi}\right)
\label{Interplay} 
\ee

\nt with $\bf M$ labelling each one of the possible models in~\eqref{Models}. $\LL_{\bf M}$ is generically encoded by~\eqref{Lagrangian}, whilst $\LL_{\rhochi}$ was already provided in~\eqref{rho-Lagrangian}. The third Lagrangian in~\eqref{Interplay} encodes fermion currents coupled to the spin-1 resonances, whereas the fourth one contains tensors of the 2nd rank made out of fermions and coupled to the resonance strength field. Their components are generically defined as
\be
\LL_{\bf M\,+\,\rho_\chi}=\frac{1}{\sqrt{2}}\,\aichi\,\cJmuuichi\left(\rhochimud-\echimud\right)\,\,+\,\,\hc
\,,
\label{Currents-L-R}
\ee
\be
 \LL^\text{mag}_{\bf M\,+\,\rho_\chi}=\frac{1}{f}\,\bichi \,\cJmunuuichi\,\rhomunuchid\,\,+\,\,\hc\,,
\label{Tensors-L-R}
\ee

\nt with an implicit summation over the index $i$ and spanning over all the possible currents and tensors that can be built upon the elementary $q$, top partner $\psi$ and elementary-top partner sector $q\psi$ and $u\psi$, \eg $i$ can denote the set $i=\{q,\,\psi,\,q\psi,\,u\psi\}$. An internal sum over the $SO(4)$ unbroken generators indices is once again implied through~\eqref{Currents-L-R}-\eqref{Tensors-L-R}. Generic coefficients $\aichi$ and $\bichi$ have been introduced and are correspondingly weighting each one of the fermion currents and tensors defined later on.

\subsection{$\fA$ and $\oA$ coupled to $\rho$}

\nt The leading order Lagrangian corresponding to $\bf{5}$-elementary fermions is given by the kinetic terms
\be
\LL_{\text{elem}}= i\,\overline{q}_L \slashed{D}\,q_L\,\,+\,\,i\,\overline{u}_R\slashed{D}\,u_R,
\label{fA-oA-elem}
\ee

\nt whereas both of the top partners $\fourplet$ and $\singlet$ are introduced in $\LL_{\text{comp}}$~\eqref{Lagrangian} through the parametrization~\cite{DeSimone:2012fs} as
\be
\begin{aligned}
\LL_{\text{comp}}= i\,\Barfourplet\slashed{\nabla}\fourplet - M_{\bf{4}}\,\Barfourplet \fourplet\,+\,\left(\fourplet \leftrightarrow\singlet\right)\,+\,\frac{f^2}{4}d^2\,+\,\left(i\,c_{41}\, (\Barfourplet)^i \gamma^\mu d_\mu^i \singlet + {\rm h.c.}\right)
\label{fA-oA-comp}
\end{aligned}
\ee

\nt with $\nabla$ standing for $\nabla=\slashed{D}+i\slashed{e}$. Goldstone bosons kinetic terms are contained at the $d^2$-term, while the coefficient $c_{41}$ controls the strength of the interplaying fourplet-singlet partner term, and it is is expected to be order one by power counting~\cite{Giudice:2007fh}. The covariant derivatives through~\eqref{fA-oA-elem}-\eqref{fA-oA-comp}, together with the $d$ and $e$-symbols are defined in~\ref{CCWZ}. Finally, the mass terms mixing the elementary and top partners are described via 
\be
\begin{aligned}
\LL_{\text{mix}} =& y_L f \left(\qBar5plet\,\rmU\right)_i \,\left(\fourpletR\right)^{i}\,\,+\,\,y_R f \left(\uBar5plet\,\rmU\right)_i \,\left(\fourpletL\right)^{i}\,\,+\,\,\hc\,\,+\,\,,\\[5mm]
& + \tilde{y}_L f \left(\qBar5plet\,\rmU\right)_5\singletR\,\,+\,\,\tilde{y}_R f \left(\uBar5plet\,\rmU\right)_5\singletL\,\,+\,\,\hc
\label{fA-oA-mix}
\end{aligned}
\ee
\begin{table}[htb!]
\centering
\small{
\hspace*{-3mm}
\renewcommand{\arraystretch}{1.0}
\begin{tabular}{c||c}
\hline\hline
\\[-3mm]
$\fA$ & $\fB$
\\[0.5mm]
\hline\hline
\\[-2mm]
$\begin{array}{l}  
\cJmuuq=\,\qBar5plet\,\,\gamma^\mu\,\BarT\,\,\5qplet\\
\\[-1mm]
\cJmuupsi =\,\Barfourplet\,\gamma^\mu\,\tau\,\,\fourplet
\\[3mm]
\cJmuuupsi =\,\left(\uBar5plet\,\BarT\,\rmU\right)_j\gamma^\mu\left(\fourpletR\right)^j
\\[3mm]
\cJmuuqpsi =\,\left(\qBar5plet\,\BarT\,\rmU\right)_j\gamma^\mu\left(\fourpletL\right)^j\\[3mm]
\end{array}$  &  
$\begin{array}{l}  
\cJmuuq =\,\left(U^T\,\q14Barplet\,U\,T\right)_{5\,j}\,\gamma^\mu\,\left(U^T\,\14qplet\,U\right)_{j\,5}\\
\\[-1mm]
\cJmuupsi =\,\Barfourplet\,\gamma^\mu\,\tau\,\,\fourplet\\[4mm]
\cJmuuqpsi =\,\left(U^T\,\q14Barplet\,U\,T\,\right)_{5\,j}\gamma^\mu\,\left(\fourpletL\right)^j
\end{array}$\\[2mm]  
\hline\\[0.5mm]
$\begin{array}{l}  
\cJmunuupsi
 =\,\BarfourpletL\,\sigma^{\mu\nu}\,\tau\,\fourpletR\\
\\[-1mm]
\cJmunuuqpsi =\,\left(\qBar5plet\,\,\BarT\,\rmU\right)_j\,\sigma^{\mu\nu}\left(\fourpletR\right)^j
\\[3mm]
\cJmunuuupsi =\,\left(\uBar5plet\,\,\BarT\,\rmU\right)_j\,\sigma^{\mu\nu}\left(\fourpletL\right)^j
\\[3mm]
\cJmunuuqu =\,\qBar5plet\,\sigma^{\mu\nu}\,\BarT\,\u5plet\\[3mm]
\end{array}$  &  
$\begin{array}{l}  
\\[-9mm]
\cJmunuupsi
 =\,\BarfourpletL\,\sigma^{\mu\nu}\,\tau\,\fourpletR\\
\\[2mm]
\cJmunuuqpsi =\,\left(U^T\,\q14Barplet\,U\,T\right)_{5\,i}\,\sigma^{\mu\nu}\left(\fourpletR\right)^i
\end{array}$\\[1mm]  
\hline\hline
\\[-3mm]
$\oA$ & $\oB$
\\[0.5mm]
\hline\hline
\\   
$\begin{array}{l}  
\cJmuuq =\,\qBar5plet\,\,\gamma^\mu\,\BarT\,\,\5qplet
\\[5mm]
\cJmuuqpsi =\,\left(\qBar5plet\,\rmU\right)\,\gamma^\mu\,\singletL\\[4mm]
\end{array}$  &  
$\begin{array}{l}  
\\[-6mm]
\cJmuuq =\,\left(U^T\,\q14Barplet\,U\,T\right)_{5\,j}\,\gamma^\mu\,\left(U^T\,\14qplet\,U\right)_{j\,5}
\end{array}$
\\[5mm]  
\hline\\[0.5mm]
$\begin{array}{l}  
\cJmunuuqpsi =\,\left(\qBar5plet\,\rmU\right)\sigma^{\mu\nu}\,\singletR
\\[5mm]
\cJmunuuqu =\,\qBar5plet\,\sigma^{\mu\nu}\,\BarT\,\u5plet\\[3mm]
\end{array}$  & 
$\begin{array}{l}  
\\[-5mm] 
-
\end{array}$\\[2mm]  
\hline \hline
\end{tabular}
\caption{\sf Currents and 2nd rank tensors for all the models. Definition $\BarT\,\equiv\,\rmU\,T\,\rmU^\dagger$ is involved through the currents, where $T$ stands as short notation for $SO(4) \simeq SU(2)_L \times SU(2)_R$ unbroken generators $T^a_\chi$ ($\chi= L,\,R$ and $a = 1,2,3$), defined in Appendix~\ref{CCWZ} together with the matrices $\tau^a$ in~\eqref{taus} have been introduced in order to keep the invariance under global $SO(5)$ transformations. Notice that the tensors are vanishing for the model $\oB$. The explicit form of the fermion currents after the vacuum alignment are reported in Appendix~\ref{Fermion currents}.}  
\label{Fermion-currents-set}
}
\end{table}

\nt Suitable $\rmU$ insertions have been done in order to guarantee the non-linear $SO(5)$ invariance. The small mixings $y_\chi$ and $\tilde{y}_\chi$ trigger the Goldstone symmetry breaking,  providing thus a proper low Higgs mass. The latter Lagrangian entails partially composite $u^5_R$ and it gives rise to quark mass terms as well as trilinear couplings contributing to the single production of top partners. Their mass spectrum, couplings, implied phenomenology, production mechanisms and relevant decay channels at LHC searches, are thoroughly analysed in~\cite{DeSimone:2012fs} for the case of a totally composite top quark $t_R$.

Altogether, the leading order composite and mixing Lagrangians contain seven parameters $\{M_{\bf{4}},\,M_{\bf{1}},\,c_{41},\, y_L,\, y_R,\, \tilde{y}_L,\, \tilde{y}_R\}$, aside from the Goldstone decay constant $f$. Six of them are arranged to reproduce the correct top mass plus the extra partner masses $\{m_{\Xft},\,m_{\Xtt},\,m_{\cT},\,m_{\cB},\,m_{\widetilde{\cT}}\}$. Their expressions are not reported here as we do not scan over them in this work.

The set of fermion currents and 2nd rank tensors constructable for both of the models $\fA$ and $\oA$ are listed in Table~\ref{Fermion-currents-set} (left column). It is worth to comment that no currents built upon elementary right handed quarks are allowed for these models as the current $\cJmuuauchi =\,\uBar5plet\,\,\gamma^\mu\,\BarT^a_\chi\,\u5plet$ turns out to be vanishing, with the definition $\BarT^a_\chi\,\equiv\,\rmU\,T^a_\chi\,\rmU^\dagger$. Parametric dependence is extended by 16 additional weighting coefficients (4 $\alpha^{L(R)}$ and 4 $\beta^{L(R)}$) at $\fA$, whilst a top partner in a singlet embedding leads to a lower number of invariants, reducing thus the parametric freedom by a half at $\oA$. Such freedom shortens even further at low energies after integrating out top partners and resonances from the physical spectrum, as some of the involved currents will vanish as well. Next sections will comment on this.

\subsection{$\fB$ and $\oB$ coupled to $\rho$}

\nt The elementary kinetic Lagrangian corresponding to this model is straightforwardly written
\be
\LL_{\text{elem}}= i\,\overline{q}_L \slashed{D}\,q_L\,,
\label{fB-oB-mass}
\ee

\nt  whereas the composite counterpart is reshuffled as
\be
\begin{aligned}
\LL_{\text{comp}}\quad\rightarrow\quad\LL_{\text{comp}}+i\,\overline{u}_R \slashed{D}\,u_R\,+\,\left(i\,c_{41}\, (\Barfourplet)^i \gamma^\mu d_\mu^i \singlet\,+\,i\,c_{4u}\, (\Barfourplet)^i \gamma^\mu d_\mu^i u_R + \hc\right),
\label{fB-oB-comp}
\end{aligned}
\ee

\nt where $\LL_{\text{comp}}$ corresponds to the strong sector Lagrangian of~\eqref{fA-oA-comp} augmented by the those terms mixing the fourplet $\fourplet$ with the singlet $\singlet$ and the totally composite $u_R$ through the coefficients $c_{41}$ and $c_{4u}$ respectively. The elementary and top partners sector are mixed via
\be
\begin{aligned}
\LL_{\text{mix}}&= y_L\,f\left(U^t\,\q14Barplet\,U\right)_{i\,5} \,\left(\fourpletR\right)^{i}\,\,+\,\,\tilde{y}_L\,f\left(U^t\,\q14Barplet\,U\right)_{5\,5} \,\singletR\,\,+\,\,y_R\,f\left(U^t\,\q14Barplet\,U\right)_{5\,5} \,\usinglet\,\,+\,\,\hc
\label{fB-oB-mix}
\end{aligned}
\ee

\nt This case also involves seven parameters $\{M_{\bf{4}},\,M_{\bf{1}},\,c_{41},\, c_{4u},\,y_L,\, y_R,\,\tilde{y}_L\}$, five of them are arranged  to reproduce the correct top mass, plus extra four partner masses as the degeneracy $m_{\Xft}=m_{\Xtt}$ is implied and also manifested at the previous two models. Notice that a direct mixing coupling $u_R$ and $\singlet$ has been removed by a field redefinition. Table~\ref{Fermion-currents-set} lists the associated fermion currents and 2nd rank tensors (right column). Notice that the element $\left(U^T\,\q14Barplet\,U\,T\right)_{i\,5}$ is zero for $i=1,...,5$, so we write $\left(U^T\,\q14Barplet\,U\,T\right)_{5\,i}$ instead, whereas the product $U^T\,\14qplet\,U$ turns out to be a symmetric matrix. On the other hand, the tensor $\cJmunuuqu=\left(U^T\,\q14Barplet\,U\,T\right)_{5\,5}\,\sigma^{\mu\nu}\,\usinglet$ is vanishing. For these models, the parametric dependence is augmented by 10 additional weighting coefficients (3 $\alpha^{L(R)}$ and 2 $\beta^{L(R)}$) at $\fB$, whereas the singlet top partner representation reduces it to just two fermion currents at $\oB$. That freedom becomes shortened after integrating out top partners and resonances at $\fB$.

\section{Integrating out the NP field content}
\label{Low-energy}

\nt In the following section the low energy effects from both of the top partners and vector resonances sectors are analysed after integrating them out from the physical spectrum, and via their corresponding EOM. In this case the UV effects are codified by the resonance mass scale $m_{\rho}$, higher above the top partners one, \ie $M_{\bf{4(1)}} < m_{\rho}$, driving thus the former sector to be integrated a priori while the latter a posteriori.

\subsection{Integrating out $\rho$}
\label{Int-out-vector-resonances}

\nt At energies $E\ll m_{\rho} \ll \Lambda$ one can integrate out the $\rho$-field by solving the EOM at lowest order in the derivative expansion. Generically, one gets from $\LL_{\rho_\chi}$ in~\eqref{rho-Lagrangian}
\be
\rhochimuu\,=\,\echimuu\,\,-\,\,\sum_{i}\,\frac{\aichi}{\sqrt{2}}\frac{\grhochi^2}{\mrhochi^2}\,\cJmuuichi
\label{rho-EOM}
\ee

\nt in such manner that the following dim-6 operators arise out at low energies from $\LL_{\bf M\,+\,\rho_\chi}$ in~\eqref{Currents-L-R}
\be
\LL_{\bf M\,+\,\rho_\chi}\,\,\xRightarrow{\rho-\text{EOM}}\,\,\,\, -\,\,\sum_{i,j} \aichi\,\ajchi\,\frac{\grhochi^2}{\mrhochi^2}\,\cO^\chi_{i\,j},\qquad\qquad \cO^\chi_{i\,j}=\cJuichi\,\cJujchi
\label{EFT-Lagrangian}
\ee

\nt where the subscripts $i$ and $j$ label different type of fermion fields in the model. The $SO(4) \simeq SU(2)_L \times SU(2)_R$ unbroken generators indices are again assumed through the vector and tensor products in~\eqref{EFT-Lagrangian}, \ie $\cJuichi\,\cJujchi\,\longrightarrow\,\cJmuuaichi\,\cJmuuajchi$. All the emerging dim-6 operators are listed in Table~\ref{dim-6} for the aforementioned models. Finally, magnetic dipole-like operators are also generated from Lagrangian~\eqref{Tensors-L-R}
\be
\begin{aligned}
 \LL_{\bf M\,+\,\rho_\chi,\,\text{mag}}\,\,\xRightarrow{\rho-\text{EOM}}\,\,\,\,\frac{1}{f}\sum_{i}\,\bichi \,\cJmunuuichi\,\emunuchimud\,\,+\,\,\hc
\end{aligned}
\label{Currents-tensors}
\ee
\begin{table}
\centering
\small{
\hspace*{-3mm}
\renewcommand{\arraystretch}{1.0}
\begin{tabular}{c||c}
\hline\hline
\\[-3mm]
$\fA$ & $\fB$
\\[0.5mm]
\hline\hline
\\[-2mm]
$\begin{aligned}
&\cO_{qq} = \frac{1}{2}\,\cJuq\,\cJuq\quad\quad
&&\cO^{(3)}_{q\psi} = \cJuq\,\cJuqpsi  \\[4mm] 
&\cO_{\psi\psi} = \frac{1}{2}\,\cJupsi\,\cJupsi\quad\quad
&&\cO^{(4)}_{q\psi} = \cJupsi\,\cJuqpsi \\[4mm] 
&\cO^{(1)}_{q\psi} = \frac{1}{2}\,\cJuqpsi\,\cJuqpsi\quad\quad
&&\blue{\cO^{(2)}_{u\psi} = \cJupsi\,\cJuupsi} \\[4mm]
&\blue{\cO^{(1)}_{u\psi} = \frac{1}{2}\,\cJuupsi\,\cJuupsi}\quad\quad
&&\blue{\cO^{(1)}_{qu\psi} = \cJuq\,\cJuupsi} \\[4mm] 
&\cO^{(2)}_{q\psi} = \cJuq\,\cJupsi\quad\quad
&&\blue{\cO^{(2)}_{qu\psi} = \cJuqpsi\,\cJuupsi}\\[4mm]
\end{aligned}
$  &  
$\begin{array}{l}
\cO_{qq} = \frac{1}{2}\,\cJuq\,\cJuq
\\[5mm]
\cO^{(2)}_{q\psi} = \cJuq\,\cJupsi\\[5mm]
\cO_{\psi\psi} = \frac{1}{2}\, \cJupsi\, \cJupsi
\\[5mm]
\blue{\cO^{(3)}_{q\psi} = \cJuq\,\cJuqpsi} \\[5mm]
\blue{\cO^{(1)}_{q\psi} = \frac{1}{2}\,\cJuqpsi\,\cJuqpsi}
\\[5mm]
\blue{\cO^{(4)}_{q\psi} = \cJupsi\,\cJuqpsi} \\[5mm]   
\end{array}
$\\[1mm]  
\hline\hline
\\[-3mm]
$\oA$ & $\oB$
\\[0.1mm]
\hline\hline
\\   
$\begin{aligned}
&\cO_{qq} = \frac{1}{2}\,\cJuq\,\cJuq \\[3mm] 
&\cO^{(1)}_{q\psi} = \frac{1}{2}\,\cJuqpsi\,\cJuqpsi\\[3mm] 
&\blue{\cO^{(2)}_{q\psi} = \cJuq\,\cJuqpsi}\\[2mm] 
\end{aligned}
$  &  
$\begin{array}{l}
\cO_{qq} = \frac{1}{2}\,\cJuq\,\cJuq
\end{array}$
\\[1mm]  
\hline \hline
\end{tabular}
\caption{\sf Dim-6 operators emerging after integrating out the vector resonances. Subscripts $q$, $\psi$ and $q\psi$ label elementary, partners fields and both of them respectively. The $SO(4)$ unbroken generators indices are implicit and for $\chi\equiv L,\,R$, through the vector products, \ie $\cJ_i\,\cJ_j\,\longrightarrow\,\cJmuuaichi\,\cJmuuajchi$. In blue are pointed out those operators vanishing by implementing the top partner's EOM as discussed in the text.}  
\label{dim-6}
}
\end{table}

\subsection{Integrating out $\Psi$}

\nt Through the EOM for the top partner fields it is possible to obtain the low energy implications for each one of the models described above. It turns out that the left/right handed chirality for the resonances are proportional to the left/right handed chirality of the elementary field, correspondingly suppressed either by the top partner mass $M_{\bf{4}}$ at the models $\fA$ and $\fB$, or by $M_{\bf{1}}$ at $\oA$ and $\oB$ (Appendix~\ref{Partners-EOM} for more details). After replacing the top partners fields according with the previous prescription, the final set of 4--fermion elementary operators obtained at low energies after decoupling the vector resonances and the top partners from the physical spectrum is given by 
\be
\cO_{u_L\,u_L} = \left(\bar{u}_L\,\gamma _{\mu }\,u_L\right)^2\,,\quad
\cO_{u_L\,d_L} =\left(\bar{u}_L\,\gamma _{\mu }\,u_L\right)\left(\bar{d}_L\,\gamma _{\mu }\,d_L\right)\,,\quad
\cO_{d_L\,d_L} = \left(\bar{d}_L\,\gamma _{\mu }\,d_L\right)^2\,,
\label{4-fermion-ops}
\ee

\nt whilst the magnetic dipole-like terms are 
\be
\cO^{mag}_\gamma = \bar{u}\,\sigma^{\mu\nu}\,u\,A_{\mu\nu}\,,\\
\qquad
\cO^{mag}_Z =\bar{u}\,\sigma^{\mu\nu}\,u\,Z_{\mu\nu}\,,\\
\qquad
\cO^{mag}_W = \bar{u}_R\,\sigma^{\mu\nu}\,d_L\,W^+_{\mu\nu}\,\,+\,\,\hc\,,
\label{Mag-dipole-ops}
\ee

\nt all them parametrizing the following effective Lagrangian
\be
\LL_{\bf M\,+\,\rho_\chi}\,\,+\,\,\LL^\text{mag}_{\bf M\,+\,\rho_\chi}\,\,\xRightarrow[\Psi-\text{EOM}]{\rho-\text{EOM}}\,\,\,\,
 -\frac{1}{f^2} \sum_{i} c_i\,\cO_i 
-\frac{1}{f} \sum_{V} c^{mag}_V\,\cO^{mag}_V
\label{Total-EFT-Lagrangian}
\ee

\nt where the Wilson coefficients $c_i$ are listed in Table~\ref{Wilson-coefficients}, with the parameter $\eta_L$ being defined in~\eqref{eta-parameters}. Magnetic dipole-type coefficients $c^{mag}_\gamma$ are left for later when dealing with EDM bounds in a forthcoming section. The indexes $i$ and $V$ run correspondingly over the operators in~\eqref{4-fermion-ops} and~\eqref{Mag-dipole-ops}. 
\begin{table}
\centering
\tiny{
\hspace*{-1cm}
\renewcommand{\arraystretch}{1.0}
\begin{tabular}{c||c}
\hline\hline
\\[-3mm]
$\fA$ & $\oA$
\\[0.5mm]
\hline\hline
\\[-2mm]
$\begin{array}{l}
c_{u_Lu_L}= \frac{1}{8} (\xi -1) \left[4 \eta _L \alpha^L_q \alpha^L_{q \psi}+\left(\alpha^L_q\right)^2+\alpha^R_q \left(4 \eta _L \alpha^R_{q \psi}+\alpha^R_q\right)\right]
\\[5mm]
c_{d_Ld_L}=\frac{1}{8} \left[-4 \eta _L \alpha^L_q \alpha^L_{q \psi}-\left(\alpha^L_q\right)^2-\alpha^R_q \left(4 \eta _L \alpha^R_{q \psi}+\alpha^R_q\right)\right]
\\[5mm]
c_{u_Ld_L}=-\frac{1}{8} (\xi -2) \left[12 \eta _L \alpha^L_q \alpha^L_{q \psi}+3 \left(\alpha^L_q\right)^2-\alpha^R_q \left(4 \eta _L \alpha^R_{q \psi}+\alpha^R_q\right)\right]
\\[5mm]
\end{array}
$  &  
$\begin{array}{l}
c_{u_Lu_L}= \frac{1}{8} (\xi -1) \left[ \left(\alpha^L _{q}\right)^2+\left(\alpha^R _{q}\right)^2\right]\\[5mm]
c_{d_Ld_L}=-\frac{1}{8} \left[\left(\alpha^L_q\right)^2+\left(\alpha^R_q\right)^2\right]
\\[5mm]
c_{u_Ld_L}=-\frac{1}{8} (\xi -2) \left[3 \left(\alpha^L_q\right)^2-\left(\alpha^R_q\right)^2\right]
\\[5mm]
\end{array}
$\\[1mm]  
\hline\hline
\\[-3mm]
$\fB$ & $\oB$
\\[0.5mm]
\hline\hline
\\   
$\begin{array}{l}
c_{u_Lu_L}= \frac{1}{8} (5 \xi -1) \left[4 \eta _L \alpha^L_q \alpha^L_{q \psi}+\left(\alpha^L_q\right)^2+\alpha^R_q \left(4 \eta _L \alpha^R_{q \psi}+\alpha^R_q\right)\right]
\\[5mm]
c_{d_Ld_L}=\frac{1}{8} (2 \xi -1) \left[4 \eta _L \alpha^L_q \alpha^L_{q \psi}+\left(\alpha^L_q\right)^2+\alpha^R_q \left(4 \eta _L \alpha^R_{q \psi}+\alpha^R_q\right)\right]
\\[5mm]
c_{u_Ld_L}=-\frac{1}{8} (7 \xi -2) \left[12 \eta _L \alpha^L_q \alpha^L_{q \psi}+3 \left(\alpha^L_q\right)^2-\alpha^R_q \left(4 \eta _L \alpha^R_{q \psi}+\alpha^R_q\right)\right]
\\[5mm]
\end{array}$
&  
$\begin{array}{l}
c_{u_Lu_L}= \frac{1}{8} (5 \xi -1) \left[\left(\alpha^L_q\right)^2+\left(\alpha^R_q\right)^2\right]
\\[5mm]
c_{d_Ld_L}=\frac{1}{8} (2 \xi -1) \left[\left(\alpha^L_q\right)^2+\left(\alpha^R_q\right)^2\right]
\\[5mm]
c_{u_Ld_L}=-\frac{1}{8} (7 \xi -2) \left[3 \left(\alpha^L_q\right)^2-\left(\alpha^R_q\right)^2\right]
\\[5mm]
\end{array}
$
\\[2mm]  
\hline \hline
\end{tabular}
\caption{\sf Corresponding Wilson coefficients $c_i$ for each one of the 4-fermion operators in~\eqref{4-fermion-ops}. All the coefficients have been expanded till linear order in $\eta$ and $\xi$ as well.}
\label{Wilson-coefficients}
}
\end{table}

As we mentioned before, the assumption of spin-1 resonances brings us a mass scale $m_\rho$ below the cut-off of the theory at $\Lambda=4\pi f$, entailing thus the coupling $1< g_\rho < 4\pi$. Likewise, the top partner mass scales $M_{\bf{4(1)}}$, assumed here such that $M_{\bf{4(1)}} < m_{\rho}$, also brings the couplings $g_{\bf{4(1)}}$. Hereinafter the linking relations
\be
\grhochi\equiv \frac{\mrhochi}{f},\qquad\qquad g_{\bf{4(1)}}\equiv\frac{M_{\bf{4(1)}}}{f}
\label{g-rho-gpartner}
\ee

\nt will be used throughout. As it is commonly argued in the literature, the ranges 500~GeV$\lesssim M_{\bf{4(1)}} \lesssim 1.5$~TeV and $1 \lesssim g_{\bf{4(1)}}\lesssim 3$ are the most favoured by concrete models (see~\cite{DeSimone:2012fs} and references therein). Finally, the strength of the elementary-composite couplings $y_{L(R)}$ and $\tilde{y}_{L(R)}$ is fixed by correctly reproducing the top quark mass. These arguments and the definitions in~\eqref{eta-parameters} lead to the estimative
\be
 0.3\lesssim \eta_{L(R)},\,\tilde{\eta}_{L(R)}\lesssim 1\,.
 \label{eta-theoretical-ranges} 
\ee

\nt Ranges that will be reconsidered later on when accounting for the 4-fermion operators effects, their underlying flavour patterns as well as phenomenological bounds that can be imposed on them in the next section. Despite the extremal value in~\eqref{eta-theoretical-ranges}, all the coefficients in Table~\ref{Wilson-coefficients} have been expanded for brevity purposes till order $\cO(\eta)$ assuming a small $\eta$. An $\xi$-expansion has been performed as well up to the linear order. As it can be seen, models $\fB$ and $\oB$ have equal Wilson coefficients, an equality no longer held at higher orders. In fact, for the latter case extra 4-fermion operators are brought in the framework in addition to those of~\eqref{4-fermion-ops} as
\be
\cO_{u_L\,u_R} =\left(\bar{u}_L\,\gamma _{\mu }\,u_L\right)\left(\bar{u}_R\,\gamma _{\mu }\,u_R\right)\,,\quad
\cO_{u_R\,d_L} =\left(\bar{u}_R\,\gamma _{\mu }\,u_R\right)\left(\bar{d}_L\,\gamma _{\mu }\,d_L\right)\,.
\label{4-fermion-ops-extra}
\ee

\nt They emerged at $\cO(\eta^3)$ and $\cO(\eta^4)$-order for $\fA$ and $\fB$ respectively, hence their Wilson coefficients turn out to be extremely suppressed and are phenomenologically negligible. Their expressions together with implied bounds are reported in Appendix~\ref{Extra-4f-operators}. At $\oA$ and $\oB$, no contributions appear for these operators, neither at any higher $\eta$-order as the involved currents are exactly vanishing after EOM.

It is worth to comment on the expressions of the currents before implementing EOM for the top partners. In particular, expressing the current $\cJmuuupsi$ from the model $\fA$ in the $SU(2)_L \times SU(2)_R$-generators space, it turns out to be
\be
\cJmuuupsiL =-\cJmuuupsiR =\,\frac{s_{\theta }}{2 \sqrt{2}}\left(\cJ^\mu_{t_R\mathcal{B}_R}-
\cJ^\mu_{X_{\text{5/3}R}t_R},\,\,i \left(\cJ^\mu_{X_{\text{5/3}R}t_R}-\cJ^\mu_{t_R\mathcal{B}_R}\right),\,\,\cJ^\mu_{\mathcal{T}_Rt_R}+\cJ^\mu_{X_{\text{2/3}R}t_R}\right)\,+\,\hc
\label{Vanishing-current}
\ee

\nt where each one of the currents above $\cJ^\mu_{\psi\phi}$ are defined in~\eqref{Generic-currents}, and the notation $(c_{\theta },s_{\theta })$ stands for $(c_{\theta },s_{\theta })\equiv(\cos\theta,\sin\theta)$ with $s_{\theta }\equiv \sin\theta\equiv v/f=\sqrt{\xi}$ . According to the EOMs in Appendix~\ref{Partners-EOM}, the RH fields $\mathcal{B}_R$ and $X_{\text{5/3}R}$ vanish, whilst the fields $\mathcal{T}_R$ and $X_{\text{2/3}R}$ are equal up to a relative negative sign. Thus, after the top partner's EOM, $\cJuupsi$ vanishes, leading the operators $\{\cO^{(1)}_{u\psi},\cO^{(2)}_{u\psi},\cO^{(1)}_{qu\psi},\cO^{(2)}_{qu\psi}\}$ to vanish as well. Similar analysis applies to the $\fB$-currents. After the corresponding top partner's EOM, $\cJuqpsi$ turns out to be completely zero, driving to zero therefore the operators $\{\cO^{(3)}_{q\psi},\cO^{(1)}_{q\psi},\cO^{(4)}_{q\psi} \}$. On the other hand, none of the $\oA$-currents, turns out to be null after EOMs, but the cancellation among different components of the currents $\cJuq$ and $\cJuqpsi$ leads the operator $\cO^{(2)}_{q\psi}$ to vanish. The latter remarks are all reflected when writing down the contribution to the dim-6 low energy operators. Notice indeed the Wilson coefficients listed in Table~\ref{Wilson-coefficients} whose contributions are coming from the non-vanishing operators.

Models $\fB$ and $\oB$ exhibit a peculiar situation for $\xi=\{1/5,\,2/7,\,1/2\}$, meaning $f\approx\{550,\,460,\,348\}$GeV. For these specific values, the coefficients $\{c_{u_L\,u_L},\,c_{u_L\,d_L},\,c_{d_L\,d_L}\}$ correspondingly vanish at both models, leaving us with no associated 4-fermion operators of the type in~\eqref{4-fermion-ops}. This feature disfavour the EWPT-preferred scenario $\xi=0.2$ in the shed light of FCNC processes. Nonetheless, bigger values are ruled out by EWPT, allowing thus non-zero contributions from such 4-fermion operators.

\subsection{Flavour and $\Delta F=2$ operators}

\nt All the SM flavour structures can be generated through flavour-breaking couplings assigned to just one chirality of SM quarks, $q_L$ or $u_R$. Corresponding scenarios will be called Right Compositeness (RC) and Left Compositeness (LC). Down-type sector also contains some flavour-breaking sources implicitly assumed, and whose effect will be reflected in the non-diagonal rotation matrices of the down-type quarks and their masses. 

In this work we will focus on flavour-symmetric scenarios, with two types of horizontal symmetries of SM up-type quarks and composite resonances, $U(3)^2$ and $U(2)^2$, the latter acting on the first two families.  Breaking of the given flavour groups will be generated by the interactions of elementary fermions with the composite sector. See Ref.~\cite{Matsedonskyi:2014iha} and references therein for a more detailed discussion on flavour patterns

\subsubsection*{Partial Compositeness}

\nt Tree-level $\Delta F =2$ processes may receive dominant contributions from the four-composite-fermion operators discussed in Section~\ref{Low-energy}. Flavour-changing 4-quark interactions are triggered after accounting for the EOM of top partners, plus elementary-composite mixings via some flavour pattern. FCNC processes are mostly constrained by the down-type quark sector, which has sizeable mixings only with $B_L$-resonances. The bilinears operators obtained in the previous section allow us to write down the $\Delta F=2$ operators as  
\be
\frac{1}{f^2}\,c_{d_L\,d_L}\,\cO_{d_L\,d_L}\quad\rightarrow\quad{1 \over f^2} \, (V_{\text{CKM} i 3}^{\dagger} V_{\text{CKM} 3 j})^2\, \kappa_{i j}^2 \, c_{d_Ld_L} [\bar d_i \gamma_{\mu} d_j] [\bar d_i \gamma^{\mu} d_j].
\label{Delta-F-ops} 
\ee
The coefficient $\kappa$ depends on the flavour pattern as follows from~\cite{Matsedonskyi:2014iha}:
\be
\begin{aligned}
&U(3)^2_{\text{LC}}\,,\quad
&&\kappa_{i j}=0\,,\\
&U(3)^2_{\text{RC}}\,\,\&\,\,U(2)^2_{\text{RC}}\,,\quad
&&\kappa_{i j}=1\,.
\label{kappa-flavour-patterns}
\end{aligned}
\ee

\nt The coefficient $c_{d_Ld_L}$ is listed in Table~\ref{Wilson-coefficients} for each model. For the bounds on the operator~\eqref{Delta-F-ops}, we take those in Ref.~\cite{Calibbi:2012at}, obtaining thus
\be
U(3)^2_{RC}: |c_{d_Ld_L}| \lesssim 5.4\times 10^{-7}_{(1,2)},\, 5.3\times 10^{-5}_{(2,3)}\,,\qquad U(2)^2_{RC}: |c_{d_Ld_L}| \lesssim 2.4\times 10^{-6}_{(1,3)}\,,
\label{Bounds-FCNC}
\ee

\nt where the subscripts $(i,j)$ stand for flavour indices of the four-fermion operator for which the constraint is obtained\footnote{In $U(3)^2$ LC this type of bound is absent, while in the other scenarios as in $U(2)^2_{\text{LC}}$, $U(2)^2_{\text{TC}}$ and $U(2)^2_{t_R\text{comp}}$ the exact constraint depends on a complex parameter (see~\cite{Matsedonskyi:2014iha,Barbieri:2012tu}).}. The previous bounds lead us to obtain the allowed parameter spaces correspondingly for each model in Fig.~\ref{Parameter-space-FCNC-bounds}, for which the most stringent bound in~\eqref{Bounds-FCNC} has been imposed. Two different cases have been considered in accordance with~\eqref{eta-theoretical-ranges}: the $\eta_L$-small limit ($\eta _L=0.3$) and the $\eta_L$-large case ($\eta _L=1$), both of them accounted for at the top and bottom respectively. The scan along $\xi=\{0.1,\,0.2,\,0.25\}$ has been performed, indicated by the \{thick, dashed, dotted\}-border inner areas respectively, and simultaneously maintaining two fermion currents on per scanning. Only three parameter spaces out of the six possibles for $\eta_L$-small limit, and three out of the fifteen available for the $\eta_L$-large case at $\fA$ have been displayed for briefness reasons. Parameter spaces provided by $\fB$ are correspondingly similar to those at $\fA$, whereas those for $\oA$ and $\oB$ are alike the one for $(\alpha^L_q,\,\alpha^R_q)$ at $\fA$, as it is inferred by their associated Wilson coefficients $c_{d_L\,d_L}$ in Table~\ref{Wilson-coefficients}. Some comments are in order:

\begin{itemize}

\item The order of magnitude of the involved coefficients turns out to be $\cO(10^{-3})$ roughly, rising till $10^{-2}$ by imposing the least stringent limit in~\eqref{Bounds-FCNC}.
\begin{figure}
\begin{center}
\includegraphics[scale=0.36]{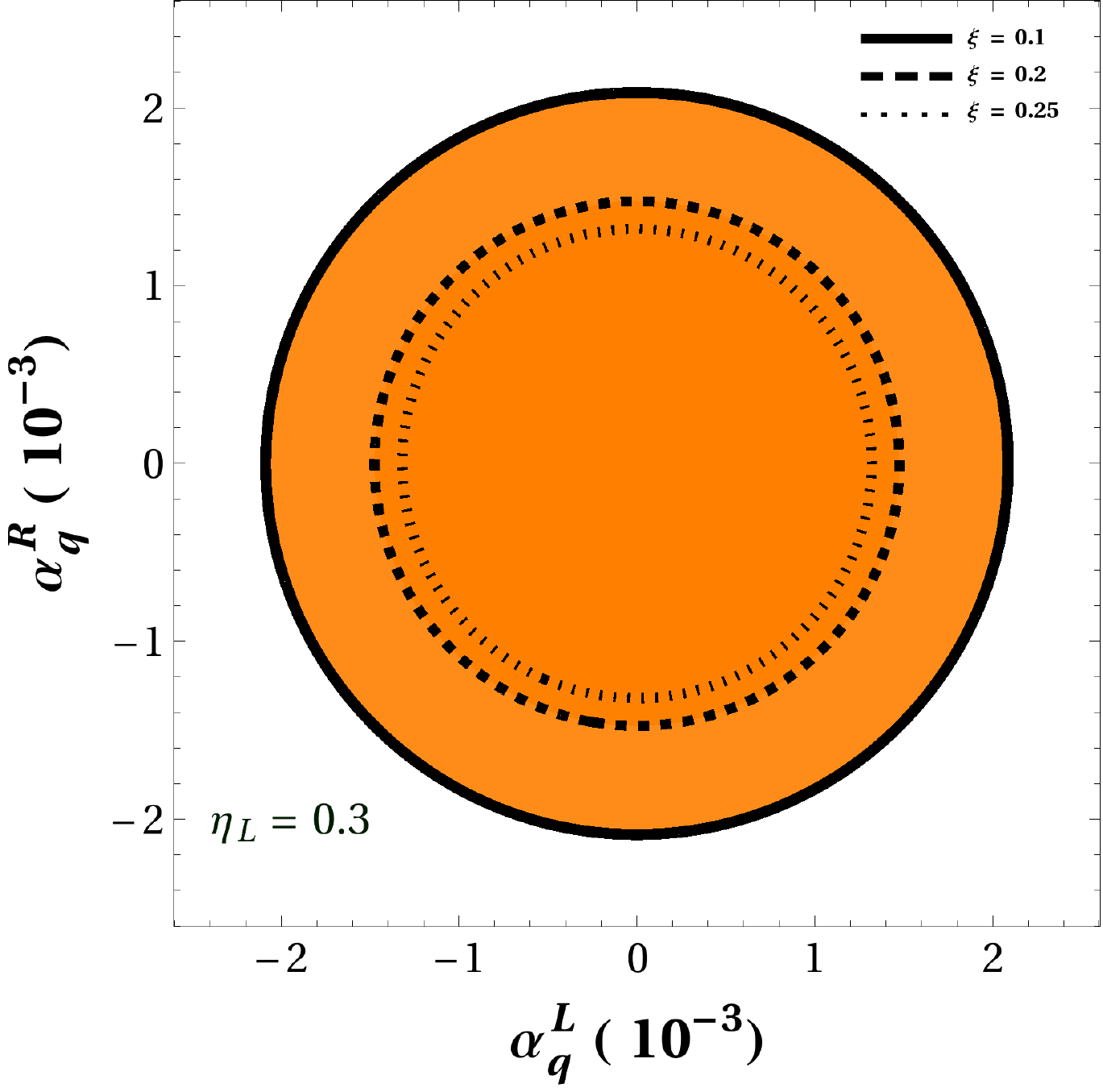}
\vspace*{1.cm}
\includegraphics[scale=0.355]{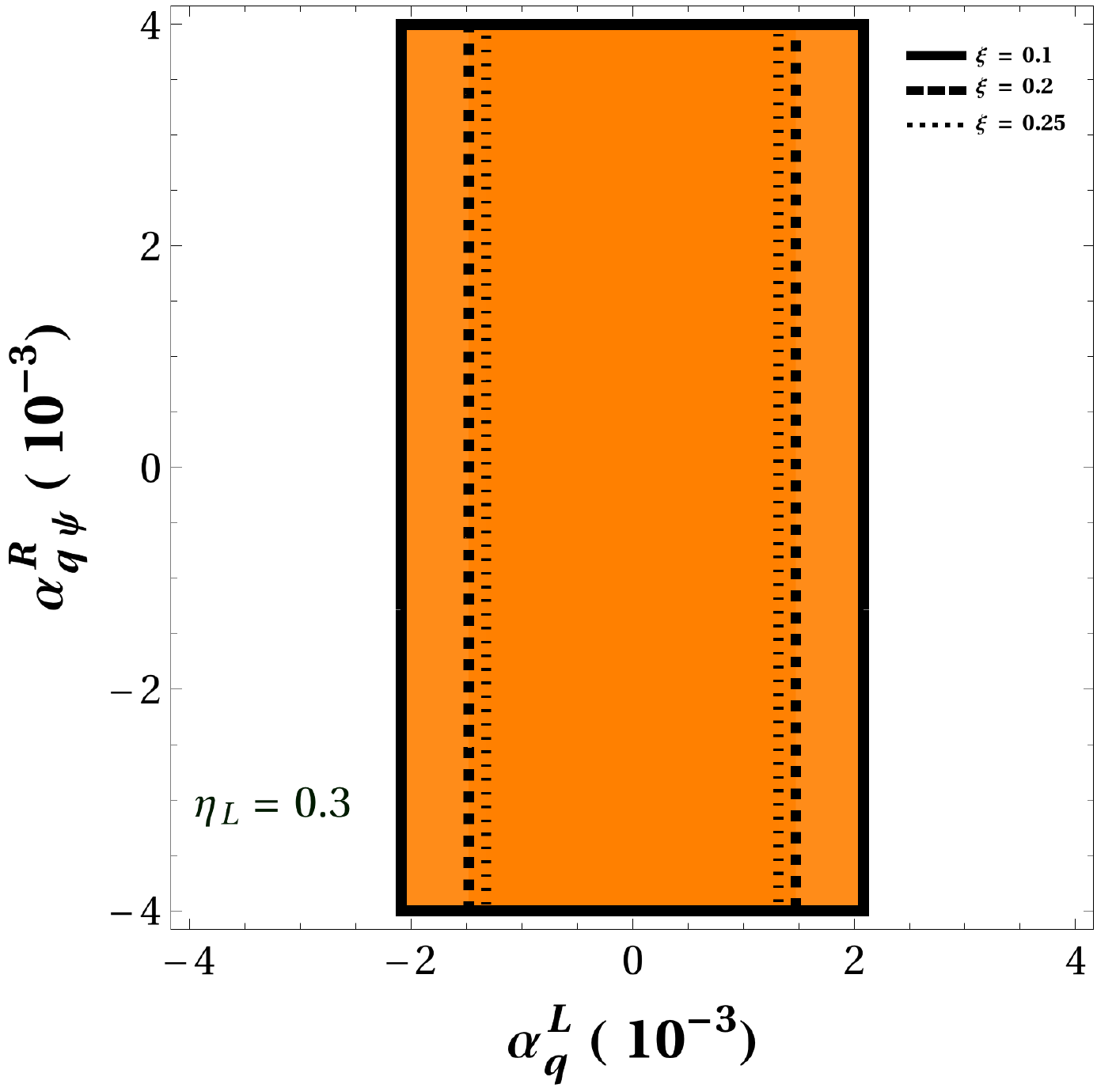}
\includegraphics[scale=0.35]{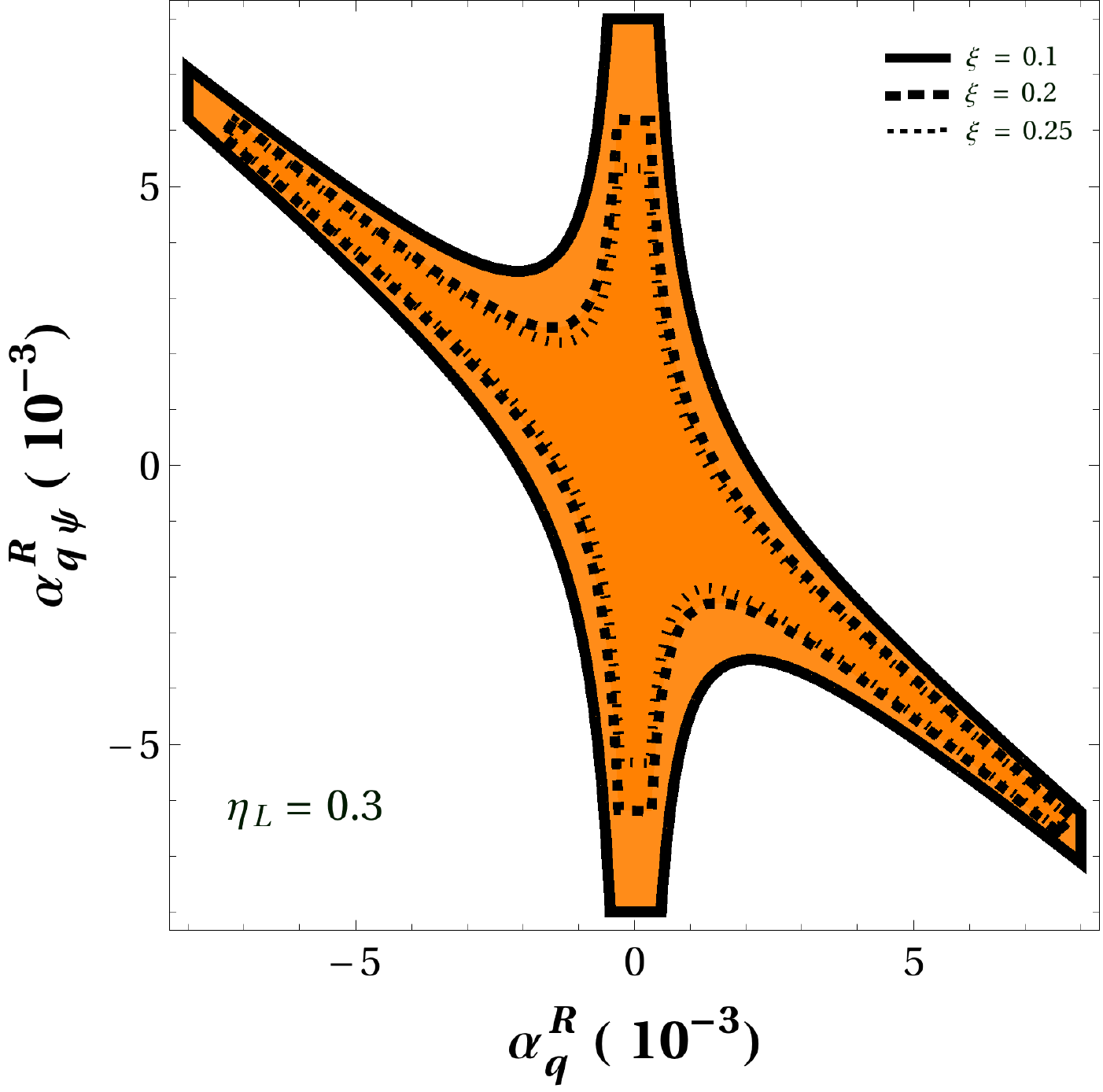}
\includegraphics[scale=0.36]{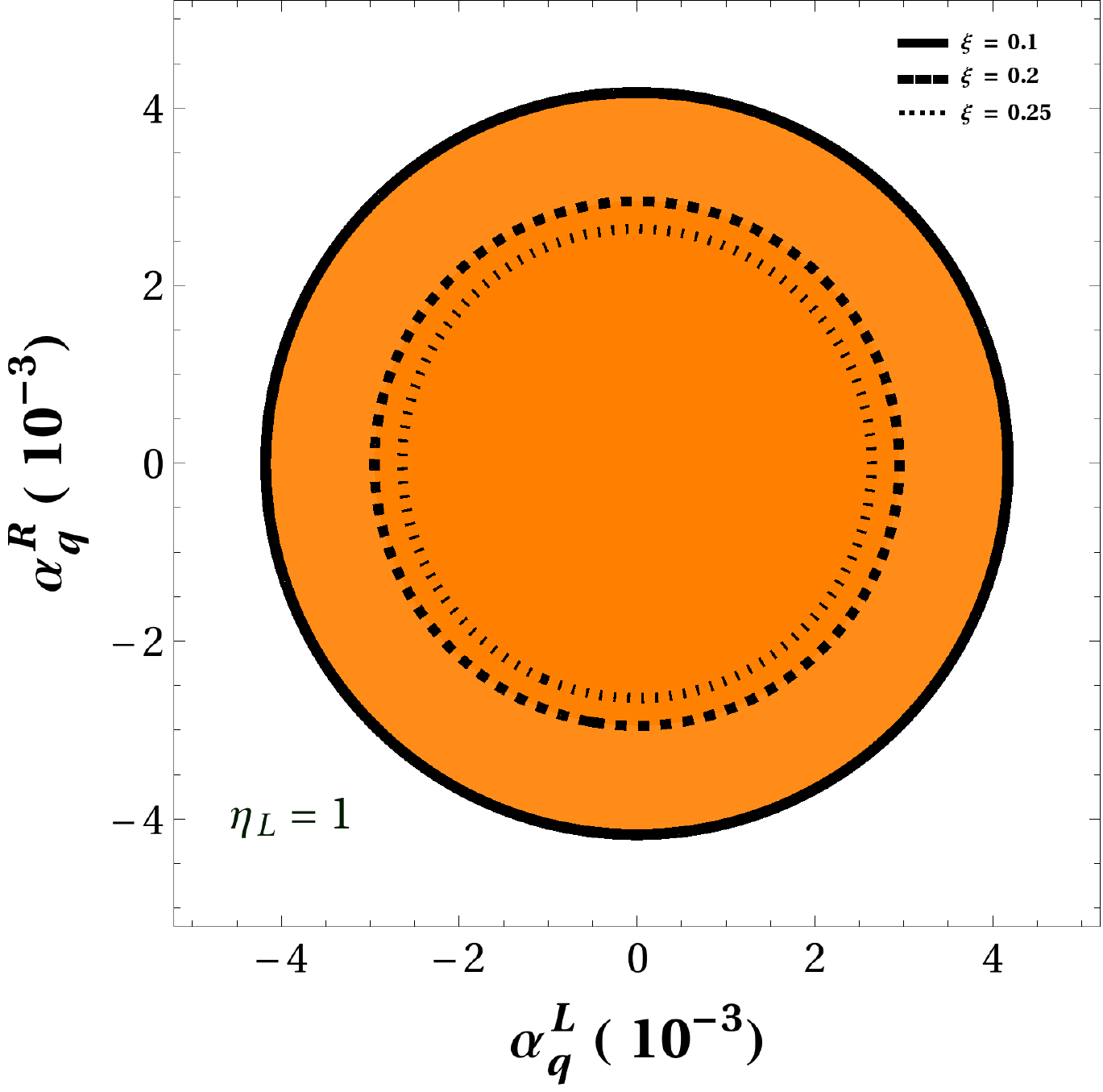}
\includegraphics[scale=0.355]{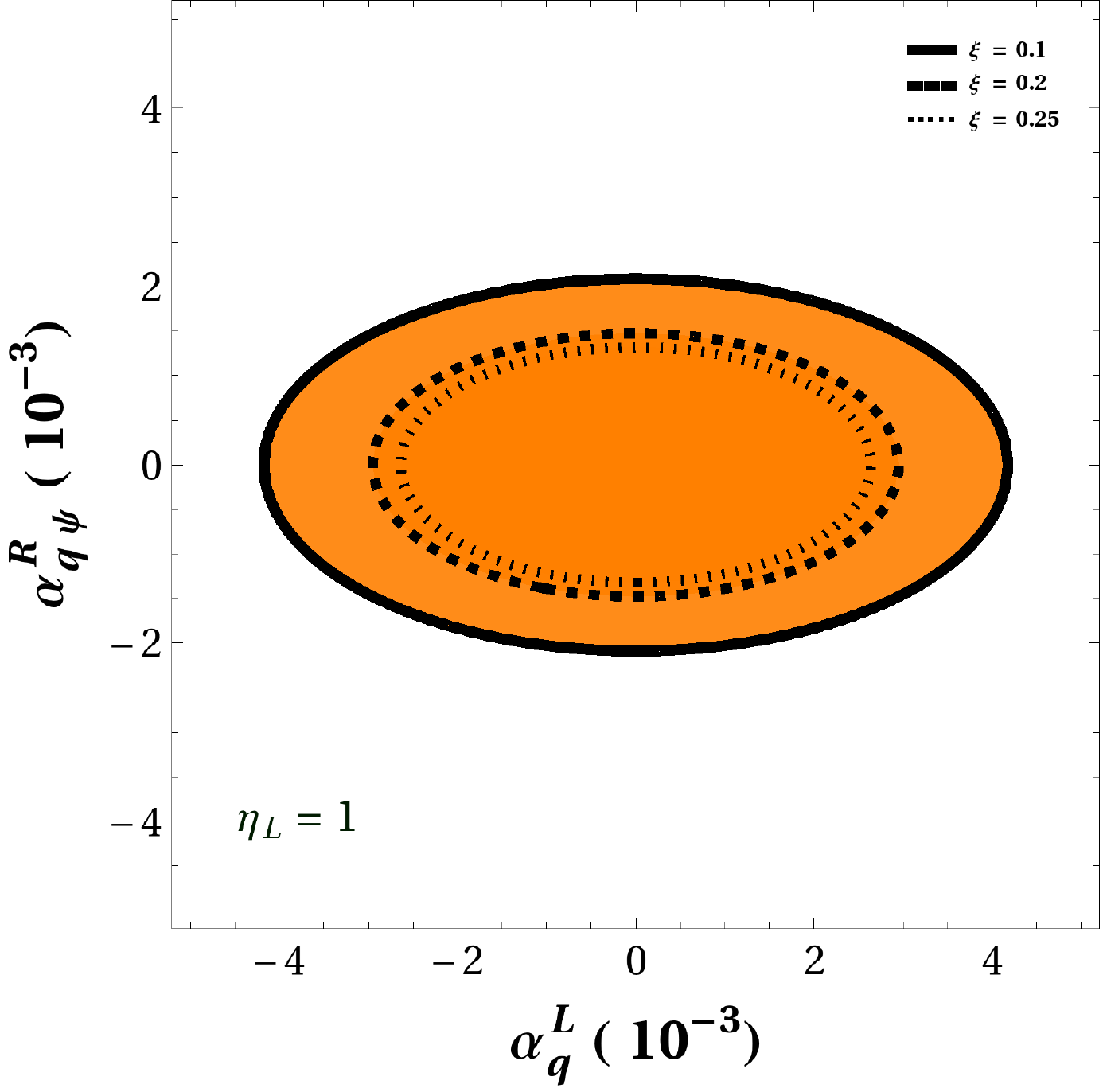}
\includegraphics[scale=0.36]{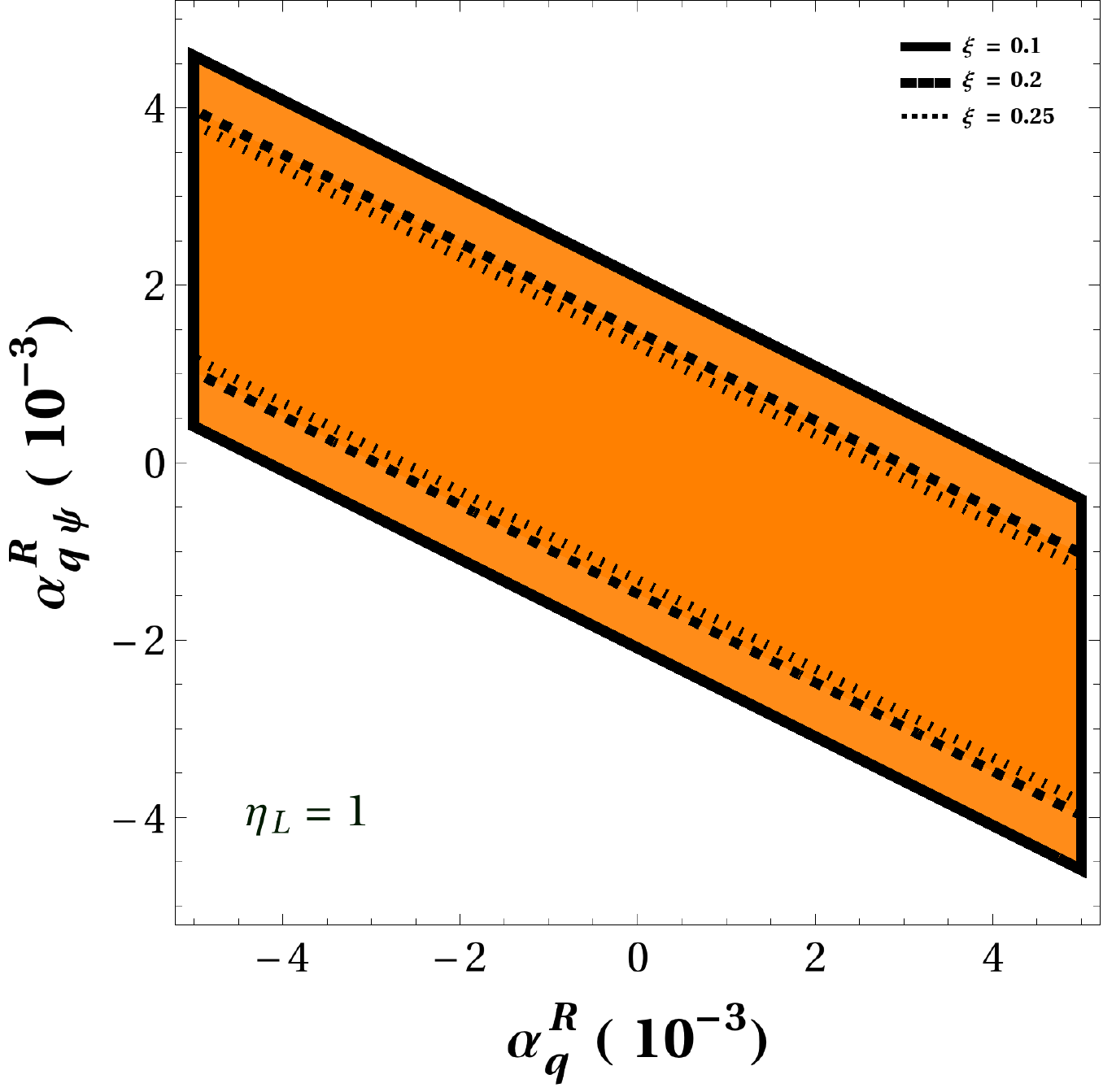}
\caption{\sf Set of parameter spaces at $\fA$ for both $\eta_L$-small and $\eta_L$-large limits (top and bottom). Those entailed by $\fB$ are correspondingly similar to those at $\fA$, while at $\oA$ and $\oB$ are similar as to the corresponding one at $\fA$ on the left, as it is inferred by their associated coefficients $c_{d_L\,d_L}$ in Table~\ref{Wilson-coefficients}. See the text for details.}
\label{Parameter-space-FCNC-bounds}
\end{center}
\end{figure}

\item One order of magnitude more is reachable in some regions. For instance, in the $\eta_L$-small limit, coefficients $\{\alpha^R_q,\,\alpha^R_{q \psi}\}$ may reach $\cO(10^{-2})$ (while $\{\alpha^L_q,\,\alpha^L_{q \psi}\}$ are off) at $\fA$ in  Fig.~\ref{Parameter-space-FCNC-bounds}, top-right. Analogous situation happens on the latter coefficients by exchanging $R \leftrightarrow L$ at the same model, whose corresponding parameter space is not shown for briefness purposes. 

\item The model $\fA$ implies certain freedom for $\alpha^R_{q\psi}$ (setting $\{\alpha^R_q,\,\alpha^L_{q\psi}\}$ to zero), inferred from the associated parameter space at the $\eta_L$-small limit in Fig.~\ref{Parameter-space-FCNC-bounds}, top-centre. Such freedom may lead to dangerous contributions to the vector resonance production through elementary-top partners collision as suggested by the current-resonance coupling in~\eqref{Currents-L-R}
\be
\aichi\,\cJuichi\,\rho^\chi \quad\longrightarrow\quad \alpha^R_{q \psi}\,\cJuqpsi^R\,\rho^R \equiv \alpha^R_{q\psi}\,\left(\qBar5plet\,\BarTR\,\rmU\gamma^\mu\fourpletL\right)\,\rho_{\mu R}\,\,+\,\,\hc\,,
\label{Dangerous-coupling}
\ee

\nt where the $SO(4)$ indices are implicitly assumed. Furthermore, dangerous contributions could be triggered for the set $\{\cO^{(1R)}_{q\psi},\,\cO^{(3R)}_{q\psi},\, \cO^{(4R)}_{q\psi}\}$. Future constraints on the $\rho$-production via elementary-top partners fusion could bound the strength of the coefficient $\alpha^R_{q\psi}$, limiting as well the aforementioned parameter space. Similar remarks, with a similar parameter space, apply by exchanging $R \leftrightarrow L$ in~\eqref{Dangerous-coupling} at the same model, and potentially affecting the operators $\{\cO^{(1L)}_{q\psi},\,\cO^{(3L)}_{q\psi},\, \cO^{(4L)}_{q\psi}\}$\footnote{Operators $\{\cO^{(2L)}_{qu\psi},\,\cO^{(2R)}_{qu\psi}\}$ are insensitive to the weighting of $\alpha^L_{q\psi}$ and $\alpha^R_{q\psi}$ as they vanish after top partner's EOM.}.

\item Likewise, the $\eta_L$-large scenario leads to a freedom, not only for $\alpha^R_{q\psi}$ but for $\alpha^R_q$ too, impacting hence on the operators $\{\cO^R_{qq},\,\cO^{(2R)}_{q\psi},\,\cO^{(3R)}_{q\psi}\}$\footnote{Other coefficients and set of operators will exhibit the same features from the others non-displayed parameter spaces.}. Analogous comments apply for $\fB$, whose coefficients $\{\alpha^R_q,\,\alpha^R_\psi\}$  exhibit similar behaviour, affecting thus the set $\{\cO^R_{qq},\,\cO^{(2R)}_{q\psi},\,\cO^R_{\psi\psi}\}$ as well as when exchanging $R \leftrightarrow L$. Such situation is sketched in Fig.~\ref{Parameter-space-FCNC-bounds} bottom-right plot for $\fA$, and it is understood by considering the Lagrangian expansion in~\eqref{EFT-Lagrangian}. Due to the quadratic $\alpha$-dependence of the Wilson coefficients, cancellations among large coefficients $\alpha^{L(R)}_i$ are possible, as long as only $SU(2)_L$ or $SU(2)_R$-currents are held on. Indeed, this brings dim-6 operators made out of mixed currents for a given $SU(2)$'s broken generator, allowing thus cancellations among opposite-sign large contributions. Conversely, holding for instance one $SU(2)_L$ and one $SU(2)_R$-currents simultaneously, will drive quadratic contributions with no mixing terms, demanding then small coefficients $\alpha$ to satisfy the imposed bounds.

\end{itemize}

\nt On the other hand, by fixing the coefficients $\alpha$ at different set of values instead, we obtain the allowed ranges for $\eta_L$ reported in Table~\ref{eta-ranges} for $\xi=0.1$. Models $\oA$ and $\oB$ provide no range as their associated Wilson coefficients are $\eta$-independent up to the order $\cO(\eta)$. Higher order terms appear when considering the $\eta_L$-large case, shaping more involved coefficients from which the ranges are inferred. Models $\oA$ and $\oB$ are disfavoured as the associated final ranges point either towards bigger negative or positive values for any $\alpha$-coefficients setting. Likewise, model $\fA$ disfavours all currents interfering either positively or negatively at both small or large limit for $\eta_L$. Different scenario occurs when a couple of currents are oppositely interfering with respect to the rest of currents (3rd column). In fact, at $\fA$ and $\fB$ the $\eta_L$-small case favours the value $\eta _L\approx 0.25$, close to the minimal one in~\eqref{eta-theoretical-ranges}. For the latter case and accordingly with~\eqref{eta-theoretical-ranges}, the intermediate value of $\eta _L\approx 0.41$ is reached at $\fA$ and $\fB$, while other ranges slightly exceed the predicted one for the same models. Different combinations of interfering currents\footnote{In the $\eta_L$-large limit at $\fA$ and $\fB$, the Wilson coefficient $c_{d_L\,d_L}$ depends on the contribution of six currents through $\{\alpha^L_q,\,\alpha^R_q,\,\alpha^L_\psi,\,\alpha^R_\psi,\,\alpha^L_{q \psi},\,\alpha^R_{q \psi}\}$. Setting them as $\alpha^L_q=\alpha^L_\psi=\alpha^R_{q \psi}=\pm 1$ and $\alpha^R_q=\alpha^R_\psi=\alpha^L_{q \psi}=\mp 1$, equal ranges as those for $\fA$ and $\fB$ at the 3rd column in Table~\ref{eta-ranges} are obtained, whilst fixing $\alpha^L_q=\alpha^R_\psi=\alpha^R_{q \psi}=\pm 1$ and $\alpha^R_q=\alpha^L_\psi=\alpha^L_{q \psi}=\mp 1$ the value $\eta _L\approx 0.42$ is obtained again.} will allow similar values and ranges as those commented for $\fA$ in the $\eta_L$-large limit (4th column). 
\begin{table}
\centering
\small{
\hspace*{-1mm}
\renewcommand{\arraystretch}{1.0}
\begin{tabular}{c||c||c||c}
\hline\hline
\multicolumn{4}{||c||}{\bf $\bf \eta_L$-small} \\[0.5mm]
\hline\hline
\bf Model & $\mathbf{\alpha _i= \pm 1}$ & $\mathbf{\alpha_i=\alpha_j = \pm 1,\, \alpha_k = \alpha_l = \mp 1}$ & \bf Other combinations
\\
\hline\hline
\\[-4mm]   
$\begin{array}{l}
\fA\\[1.5mm]
\quad\&\\[1.5mm]
\fB\\[2mm]   
\end{array}$
 &  
$\eta _L\approx -0.25$
&  
$\eta _L\approx 0.25$
& 
$-$\\[2mm]   
\hline\hline 
\multicolumn{4}{||c||}{\bf $\bf \eta_L$-large}\\[0.5mm]
\hline\hline
\\[-4mm]
\bf Model & $\mathbf{\alpha _i= \pm 1}$ 
& 
$\begin{array}{l}
\mathbf{\alpha_i=\dots= \alpha_l= \pm 1}\\[1.5mm]
\mathbf{\alpha_m=\alpha_n  = \mp 1}   
\end{array}$
& 
\bf Other combinations\\[0.5mm]
\hline\hline
\\[-4mm]   
$\begin{array}{l}
\\[1.mm]
\fA\\[1.mm]
\quad\&\\[1.mm]
\fB\\[1mm]   
\end{array}$
 &  
$\begin{array}{l}
\\[1.5mm]
-1.06\leq \eta _L\leq -0.94\\[1.5mm]
\hspace*{-1mm}\left(-1.19\lesssim \eta _L\lesssim -0.84\right)
\end{array}$
& 
$\begin{array}{l}
0.94\lesssim \eta _L\lesssim 1.06\\[1.5mm]
\hspace*{-1mm}\left(0.84\lesssim \eta _L\lesssim 1.19\right)
\end{array}$
&
$\begin{array}{l}
0.94\lesssim \eta _L\lesssim 1.06\\[1.5mm]
\hspace*{-1mm}\left(0.84\lesssim \eta _L\lesssim 1.18\right)
\end{array}$
\\[1mm] 
\cline{3-4}\\[-4mm] 
&& $\eta _L\approx 0.41$ & $\eta _L\approx 0.41$
\\[1mm] 
\hline\hline\\[-4mm] 
$\begin{array}{l}
\oA\\[1.5mm]
\quad\&\\[1.5mm]
\oB\\[2mm]   
\end{array}$
&  
$\begin{array}{l}
|\eta_L|\gtrsim 26.01\\[1.5mm]
\hspace*{-1mm}\left(|\eta_L|\gtrsim 8.21\right)\\[2mm]   
\end{array}$
&  
$\begin{array}{l}
|\eta_L|\gtrsim 26.01\\[1.5mm]
\hspace*{-1mm}\left(|\eta_L|\gtrsim 8.21\right)\\[2mm]   
\end{array}$
&
$\begin{array}{l}
|\eta_L|\gtrsim 26.01\\[1.5mm]
\hspace*{-1mm}\left(|\eta_L|\gtrsim 8.21\right)\\[2mm]   
\end{array}$
\\[4mm] 
\hline \hline
\end{tabular}
\caption{\sf Allowed $\eta_L$-ranges from bounds in~\eqref{Bounds-FCNC} for $\xi=0.1$, either in the small or large limit for the $\eta$-parameters, and by setting the $\alpha$-coefficients at different values: $\pm 1$ all at once, $\pm 1$ by pairs while the rest at $\mp 1$, and several other combinations. See the text for details. Some cases lead to no ranges as their implied Wilson coefficient becomes constant.}  
\label{eta-ranges}
}
\end{table}
Following previous discussion and according to definitions in~\ref{eta-parameters}, some rough relations among the Yukawa $y_L$ and the coupling $g_{\bf{4}}$ may be derived. Indeed, the beforehand estimation in~\eqref{eta-theoretical-ranges} can be translated into
\be
 0.3\,g_{\bf{4}}\lesssim y_{L(R)}\lesssim g_{\bf{4}},\qquad\qquad  0.3\,g_{\bf{1}}\lesssim \tilde{y}_{L(R)}\lesssim g_{\bf{1}}\,.
 \label{eta-theoretical-ranges-y-g} 
\ee

\nt Ranges in Table~\ref{eta-ranges} point to $y_L\approx 0.25(0.41)\,g_{\bf{4}}$ at $\fA$ and $\fB$. Estimations for $y_R$ and $\tilde{y}_{L(R)}$ are derived once the Wilson coefficient are expanded up to the order $\cO(\eta^2)$ or higher. See Appendix~\ref{Extra-4f-operators} for more details.

\subsubsection*{Compositeness constraints}

\nt Concerning flavour patterns, where the degree of light quark compositeness is related to the one of the top quark, stringent constraints come from the searches for quark compositeness. 
Among the 4-fermion operators up and down quarks in LC, one of the most constraining is $\cO^{(1)}_{qq}=(\bar{q}_L\gamma^\mu q_L)(\bar{q}_L\gamma_\mu q_L)$, tradable in terms of the subset $\{\cO_{u_L\,u_L},\,\cO_{u_L\,d_L},\,\cO_{d_L\,d_L}\}$ in~\eqref{4-fermion-ops} via the following equivalence among operator coefficients at the Lagrangian level
\be
c_{u_L\,u_L}=2\,c_{u_L\,d_L}=c_{d_L\,d_L}=c^{(1)}_{qq}\,.
\label{Coeff-relations}
\ee

\nt In addition to $\cO^{(1)}_{qq}$, one 4-fermion operator in both LC and RC scenario are disregarded here, as the former requires colour effects not included in this work, whereas the latter contains right handed up quarks only, not emerging in our framework (see~\cite{Barbieri:2012tu} for more details). $\cO^{(1)}_{qq}$ gives rise to departures in the jet's angular distributions with respect to the SM predictions, therefore when $q_L$ belongs to the first quarks generation the corresponding Wilson coefficient is bounded as $(5.0\,\text{TeV})^{-2}$~\cite{Domenech:2012ai}, which for $\xi=\{0.1,\,0.2,\,0.25\}$ translates correspondingly as $c^{(1)}_{qq} \lesssim \{1.2,\,0.6,\,0.5\}\times 10^{-8}$, and consequently into
\be
c^{(1)}_{qq}  \quad\Longrightarrow\quad\left\{ \begin{array}{l}
c_{u_L\,u_L}\,\,\,\&\,\,\,c_{d_L\,d_L}\lesssim \{1.2,\,0.6,\,0.5\}\times 10^{-8}\,,\\ \\
\phantom{c_{u_L\,u_L}\,\,\,\&\,\,\,}c_{u_L\,d_L} \lesssim \{0.6,\,0.3,\,0.25\}\times 10^{-8}\,.
\end{array}
\right.
\label{Coeff-dijet-bounds}
\ee
\begin{figure}
\begin{center}
\includegraphics[scale=0.415]{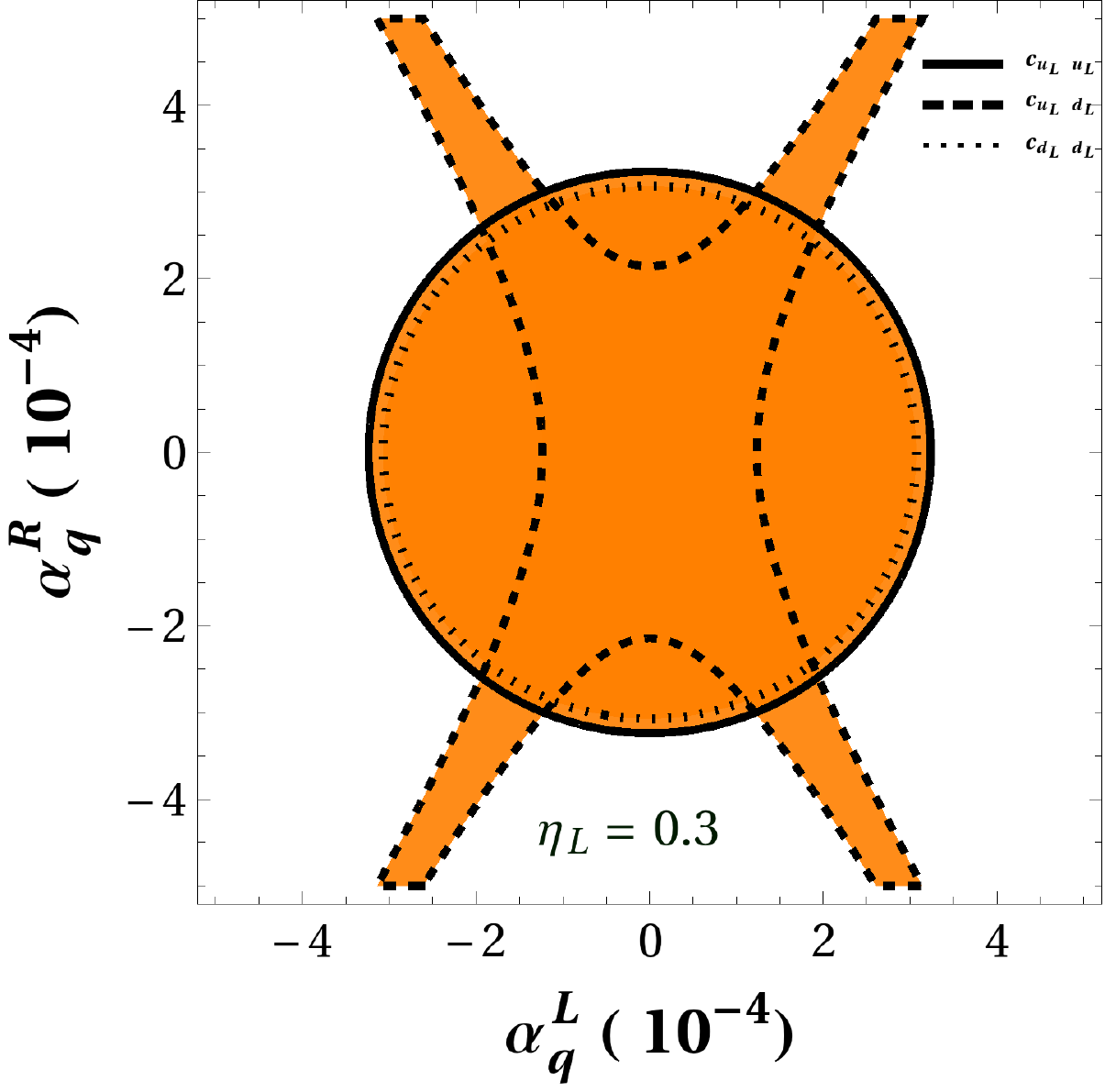}
\includegraphics[scale=0.415]{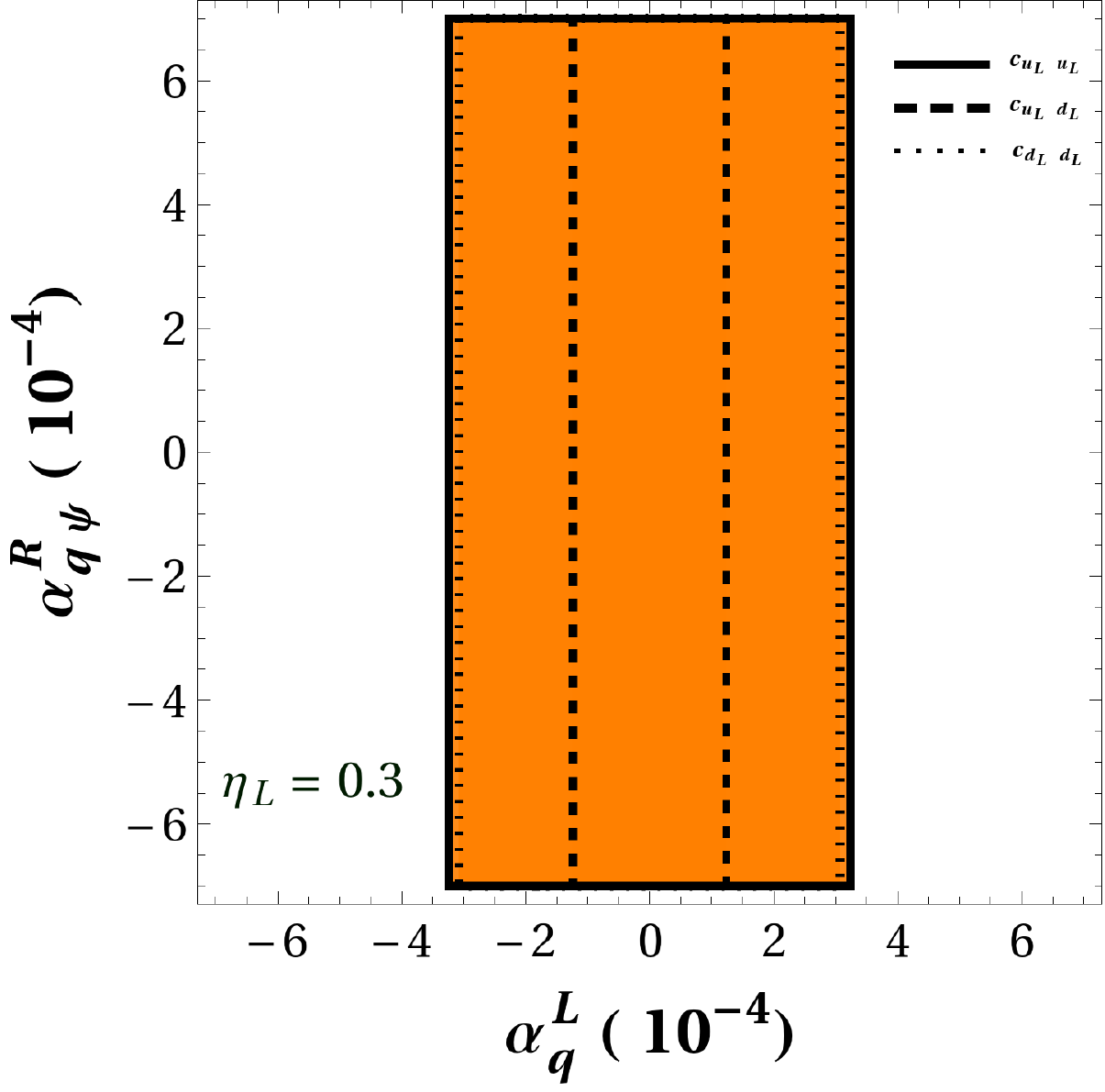}
\includegraphics[scale=0.415]{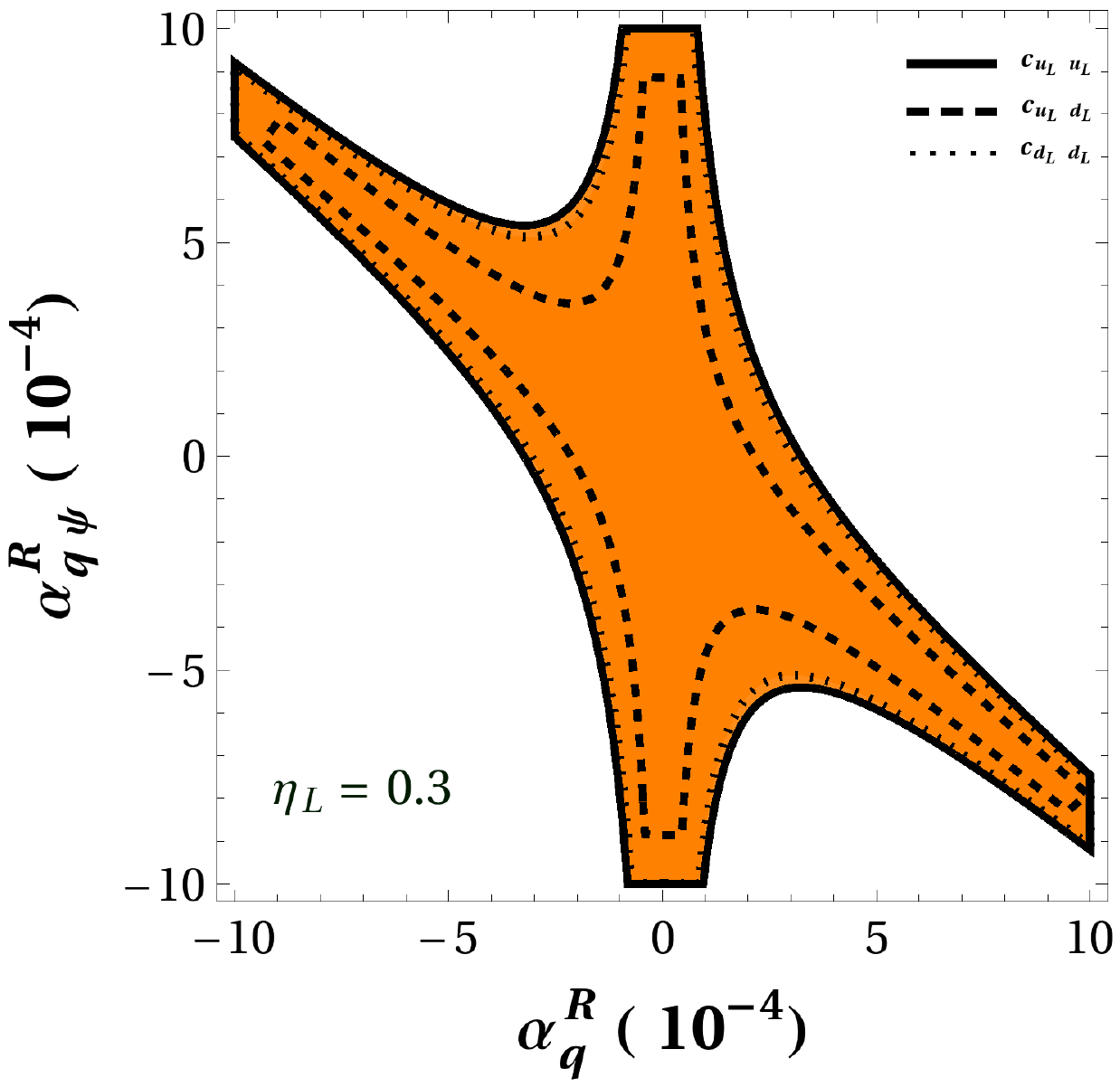}
\caption{\sf Same parameter spaces for $\fA$ as in Fig.~\ref{Parameter-space-FCNC-bounds}, but for $\eta_L=0.3$ and $\xi=0.1$ only. Parameter spaces entailed by $\fB$ are similar to those at $\fA$ respectively, whilst $\oA$ and $\oB$ are alike  the corresponding one at $\fA$ (left), as it is inferred by its associated Wilson coefficients in the limit $\alpha^{L(R)}_{q \psi}\to 0$ (see Table~\ref{Wilson-coefficients}). The \{thick, dashed, dotted\}-border inner areas corresponds to the $\{c_{u_L\,u_L},\,c_{u_L\,d_L},\,c_{d_L\,d_L}\}$-dijet bounds respectively in~\eqref{Coeff-dijet-bounds}. See text for details.}
\label{Parameter-space-Dijet-bounds}
\end{center}
\end{figure}

\nt The parameter spaces obtained by these bounds are gathered in Fig.~\ref{Parameter-space-Dijet-bounds} for the $\eta_L$-small limit and $\xi=0.1$ only. Models $\fB$ involves similar to those at $\fA$ respectively, while $\oA$ and $\oB$ are alike  the corresponding one at $\fA$ (left), as it is inferred by its associated Wilson coefficients in the limit $\alpha^{L(R)}_{q \psi}\to 0$ (see Table~\ref{Wilson-coefficients}). The three different dijet bounds in~\eqref{Coeff-dijet-bounds} are imposed by holding one operator on at each time correspondingly, being indicated in \{thick, dashed, dotted\}-border inner areas from $\{c_{u_L\,u_L},\,c_{u_L\,d_L},\,c_{d_L\,d_L}\}$-bounds respectively. Generically concluding, the coefficients turns out to be one order of magnitude smaller and more constrained with respect to those inferred by $\Delta F =2$ processes in Fig.~\ref{Parameter-space-FCNC-bounds}. The bound on $c_{u_L\,d_L}$ aims at further restricting the allowed parameter spaces to smaller regions, due to the relative factor in~\eqref{Coeff-relations}. Finally, limits for the parameter $\eta_L$ may be inferred once again, turning out to be similar as the ones reported in Table~\ref{eta-ranges} by implementing either $c_{u_Lu_L}$ or $c_{d_Ld_L}$. The case of $c_{u_Ld_L}$ will require slightly different ranges, and similar remarks as those inherent to the ranges in Table~\ref{eta-ranges} apply also.

\subsection{Fermionic EDMs bounds}
\label{subsubsec:edm}

\begin{figure}
\begin{center}
\includegraphics[scale=0.6]{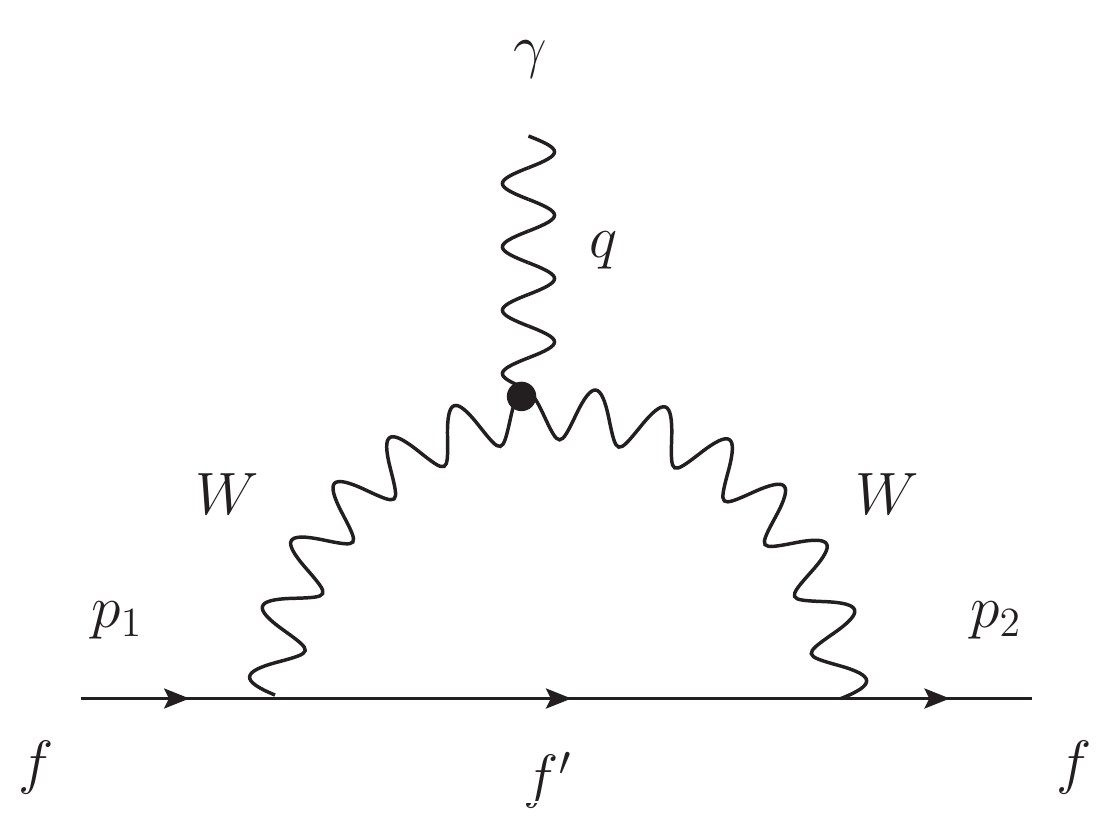}
\hspace*{0.25cm}
\includegraphics[scale=0.4]{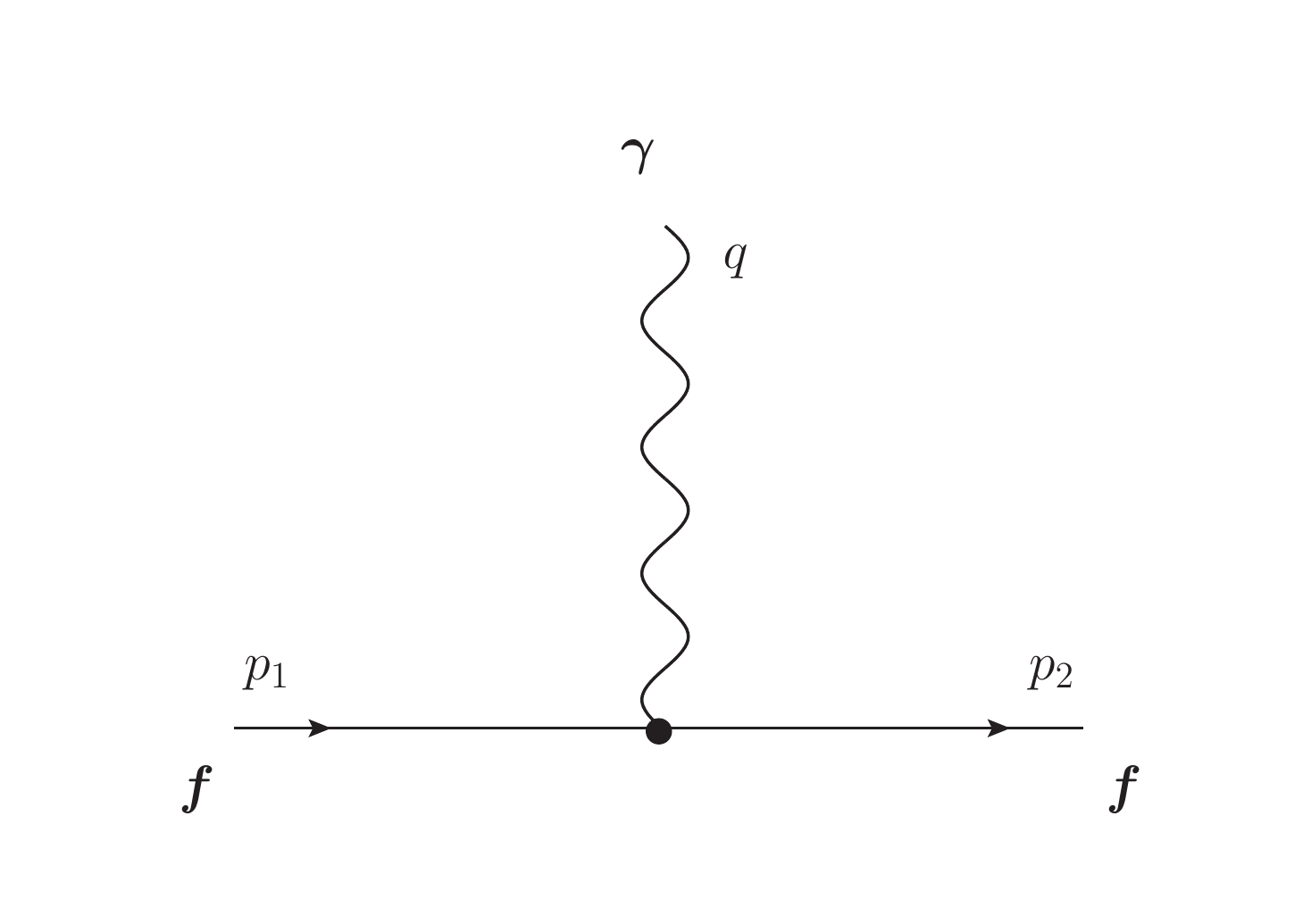}
\caption{\sf Left: 1-loop fermion EDM induced by a CP-odd TGV coupling $\frac{i}{2} \epsilon_{\mu\nu\rho\sigma }W^+_{\mu } W^-_{\nu } A^{\rho\sigma}$. Right: Effective fermionic EDM from operator $\cO^{mag}_\gamma$ in~\eqref{Mag-dipole-ops}, yielded after integrating out vector resonances and quark partners.}
\label{loop-effective-edm}
\end{center}
\end{figure}

\nt Electric dipole moments for quarks and leptons are generically
signalling BSM sources of CP-violation, as they tend to be almost free from SM background contributions as well as experimentally quite bounded. Fermionic EDMs are suppressed in the SM beyond two electroweak boson exchange, while one-loop level induced in most BSM theories. Indeed, via the effective gauge boson-photon coupling
$\frac{i}{2} \epsilon_{\mu\nu\rho\sigma }W^+_{\mu } W^-_{\nu } A^{\rho\sigma}$, a one-loop contribution for the fermion EDM is triggered, see Fig.~\ref{loop-effective-edm} (left). Such coupling naturally emerges in BSM scenarios and it is  effectively provided by non-linear operators in~\cite{Gavela:2014vra,Yepes:2015qwa,Yepes:2015zoa,Alonso:2012pz,
Alonso:2012px}.

The amplitude corresponding to the Feynman diagram in Fig.~\ref{loop-effective-edm} left, can be parametrised as
\begin{align}
\mathcal{A}_{f}&\equiv -i\,d_f\,\overline{u}\left(p_2\right)\,\sigma_{\mu \nu}q^{\nu }\gamma^5 \,u\left(p_1\right)\,,
\end{align}

\nt where $d_f$ denotes the fermionic EDM strength. Such amplitude is  effectively provided by the operator $\cO^{mag}_\gamma$ in~\eqref{Mag-dipole-ops}, as the vector resonances and quark partners are integrated out from the physical spectrum, see Fig.~\ref{loop-effective-edm} right. From \eqref{Mag-dipole-ops}-\eqref{Total-EFT-Lagrangian} we have
\be
d_f=\frac{c^{mag}_\gamma}{f}\,.
\ee

\nt Accounting for the form of $c^{mag}_\gamma$ in each one of the above models, we obtain the following expressions
\be
\begin{aligned}
&\fA:
&&\quad\quad\left|\frac{d_f}{e}\right|=\left|\frac{\sqrt{\xi }}{4\sqrt{2}f}
\left[\eta _R \left(\beta^L_{q \psi}-\beta^R_{q \psi}\right)+\beta^L _{q u}+\eta _L \left(\beta^L_{u \psi}-\beta^R_{u \psi}\right)-\beta^R_{q u}\right]\right|\,,\\ \\
&\fB:
&&\quad\quad\left|\frac{d_f}{e}\right|=\left|\frac{1}{4f}\sqrt{\frac{\xi}{2}}\,\eta _R\left(\beta^L_{q \psi}-\beta^R_{q \psi}\right)\right|\,,\\ \\
&\oA:
&&\quad\quad\left|\frac{d_f}{e}\right|=\left|\frac{\sqrt{\xi }}{4\sqrt{2}f} \left(\beta^L_{q u}-\beta^R_{q u}\right) + i\,\frac{ (\xi -2)}{4 \sqrt{2} f}\,\tilde{\eta }_R\left(\beta^L_{q \psi}+\beta^R_{q \psi}\right)\right|\,.
\label{EDM-fA-fB-oA}
\end{aligned}
\ee

\nt Associated EDM coefficients have been expanded till linear order for $\fA$ and $\fB$ in the $\eta_{L(R)}$-small limit, assuming $\tilde{\eta}_{L(R)}$ small as well. By setting values for the parameters and for the weighting coefficients in each model, one has the following phenomenological bounds.

\subsubsection*{Phenomenological bounds}
\label{subsubsec:pheno-edm}

\nt Using as values for the constituent quark masses $m_u=m_d=m_N/3$, the experimental limit on the neutron EDM~\cite{Baker:2006ts},
\be
\left|\frac{d_n}{e}\right|<2.9\times 10^{-26}\,\text{cm}\,,\qquad
\text{at 90\% CL}\,,
\label{eq:eedm}
\ee
allows to set the following strong limits on the combination of the involved coefficients by fixing the parameters as $\eta _L=\eta _R=\tilde{\eta }_L=\tilde{\eta }_R=0.3$\footnote{Such assumption naturally arises in deconstructed CHM~\cite{Panico:2011pw,DeCurtis:2011yx}, allowing to decrease the sensitivity of the Higgs potential to the cutoff scale.}
\be
\begin{aligned}
&\fA:
&&\quad\quad\Big|\xi  \left(\beta^L _{q u}-\beta^R_{q u}+0.3 \left(\beta^L_{q \psi}-\beta^R_{q \psi}\right)+0.3 \left(\beta^L_{u \psi}-\beta^R_{u \psi}\right)\right)\Big|<2.1\times 10^{-5}
\,,\\ \\
&\fB:
&&\quad\quad\phantom{\hspace*{6.5cm}}\Big|\xi  \left(\beta^L_{q \psi}-\beta^R_{q \psi}\right)\Big|<7.1\times 10^{-5}\,,\\ \\
&\oA:
&&\quad\quad\phantom{\hspace*{0.5cm}}\Big|\sqrt{\xi } \left[\sqrt{\xi }\left(\beta^L_{q u}-\beta^R_{q u}\right)+0.3\,i(\xi -2)\left(\beta^L_{q \psi}+\beta^R_{q \psi}\right)\right]\Big|<2.1\times 10^{-5}\,.
\label{Numerical-edm-bound-fA-fB-oA}
\end{aligned}
\ee

\begin{figure}
\begin{center}
\includegraphics[scale=0.43]{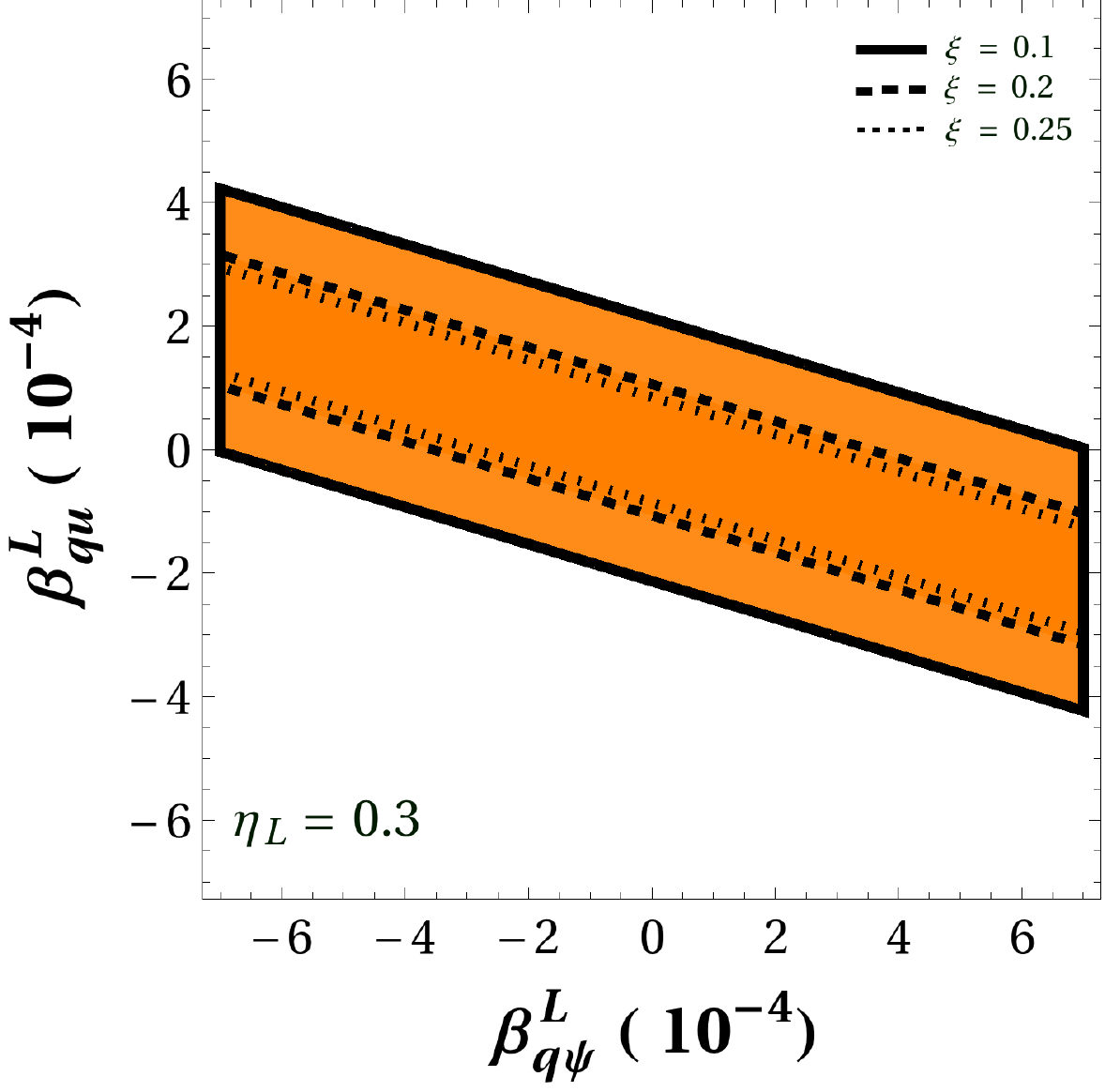}
\includegraphics[scale=0.43]{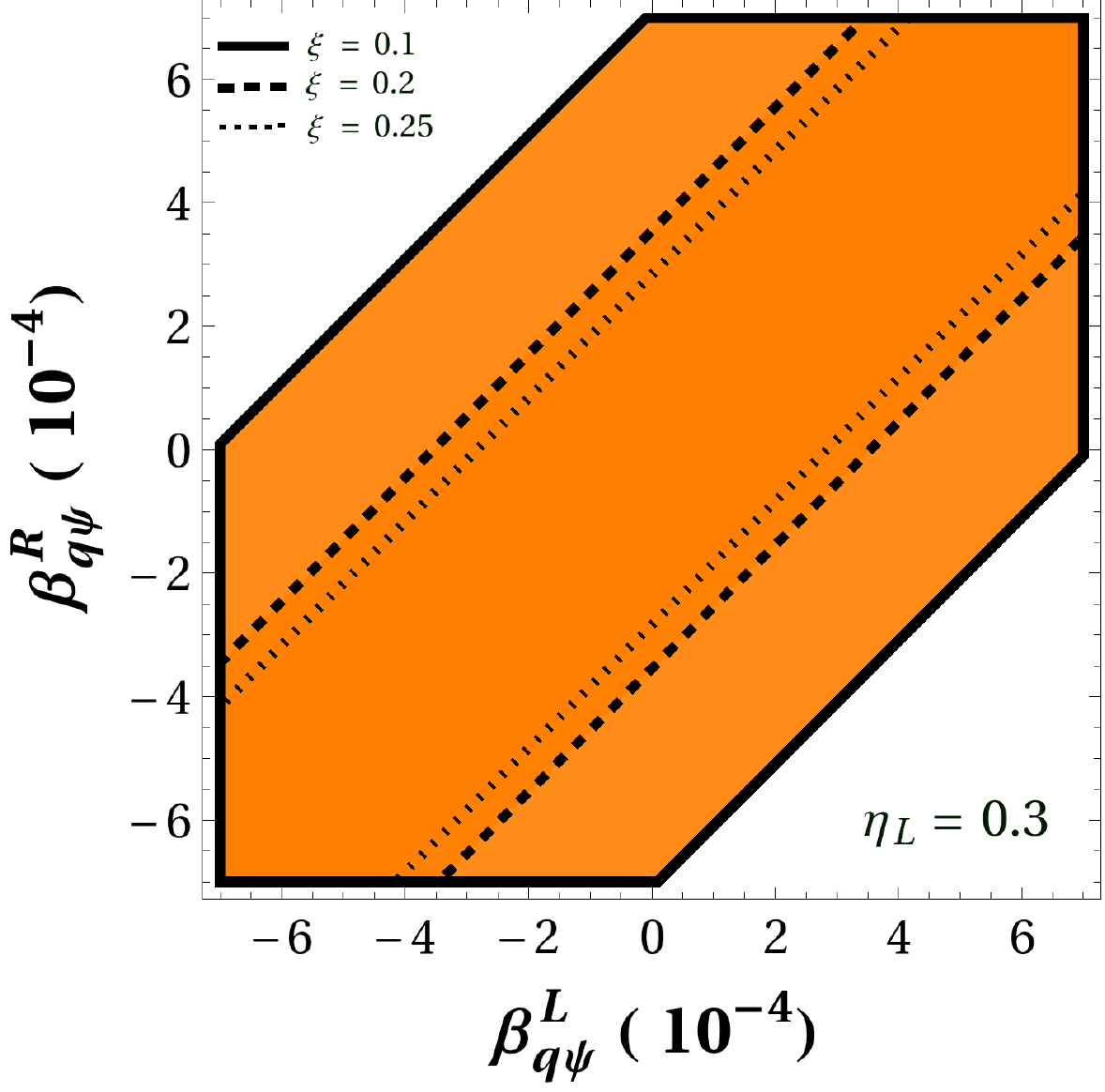}
\vspace*{1.cm}
\includegraphics[scale=0.43]{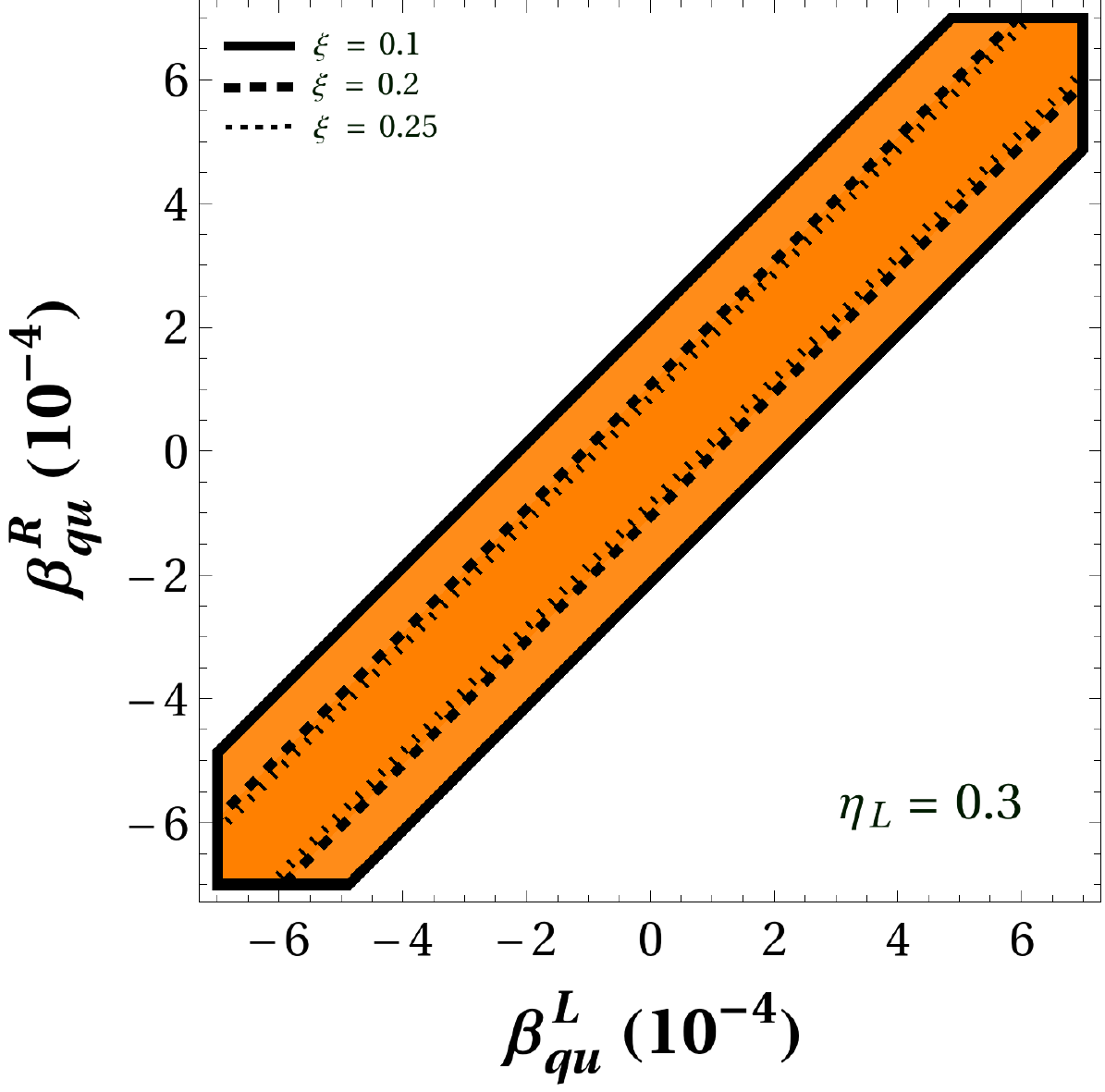}
\includegraphics[scale=0.43]{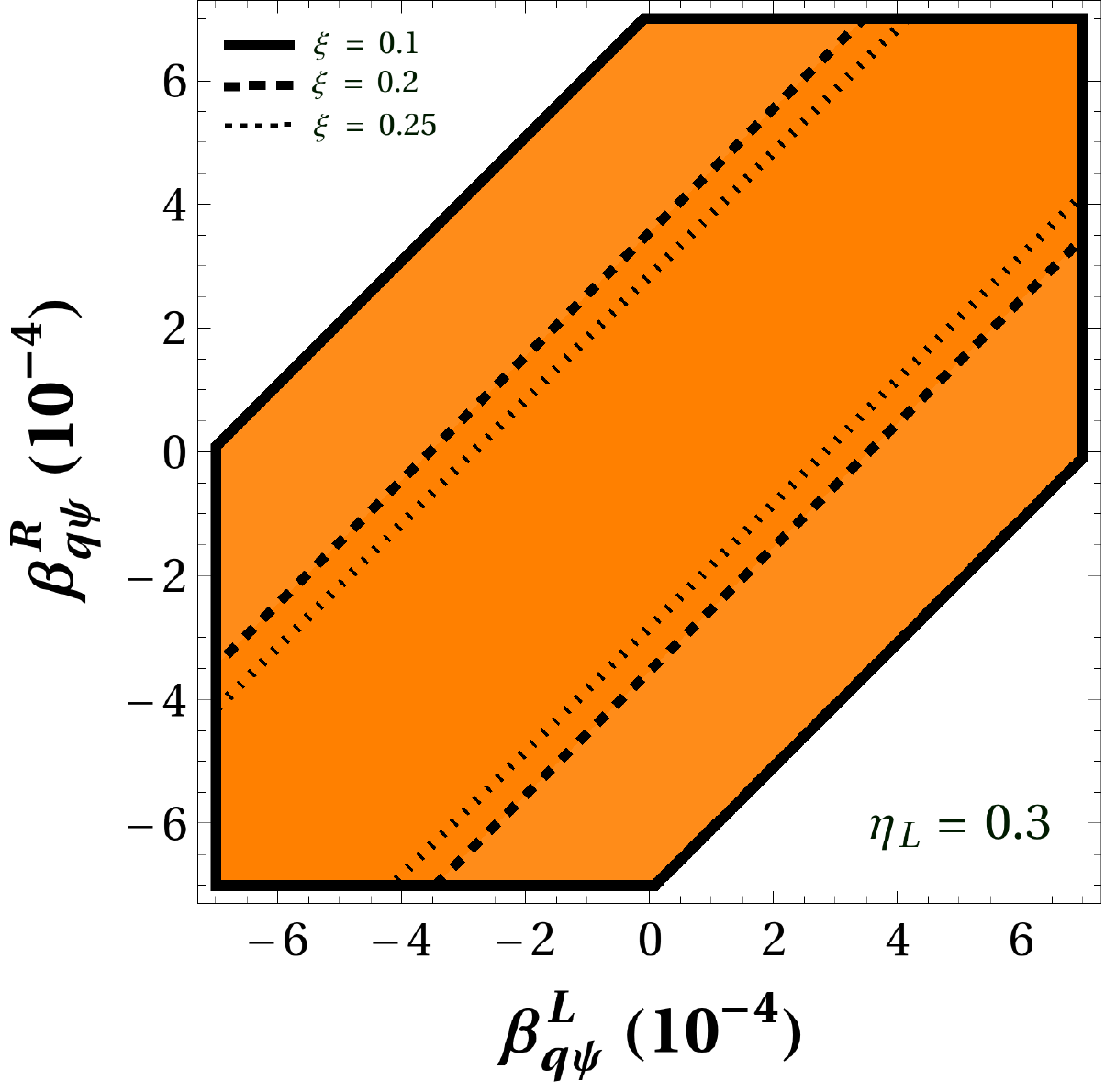}
\includegraphics[scale=0.43]{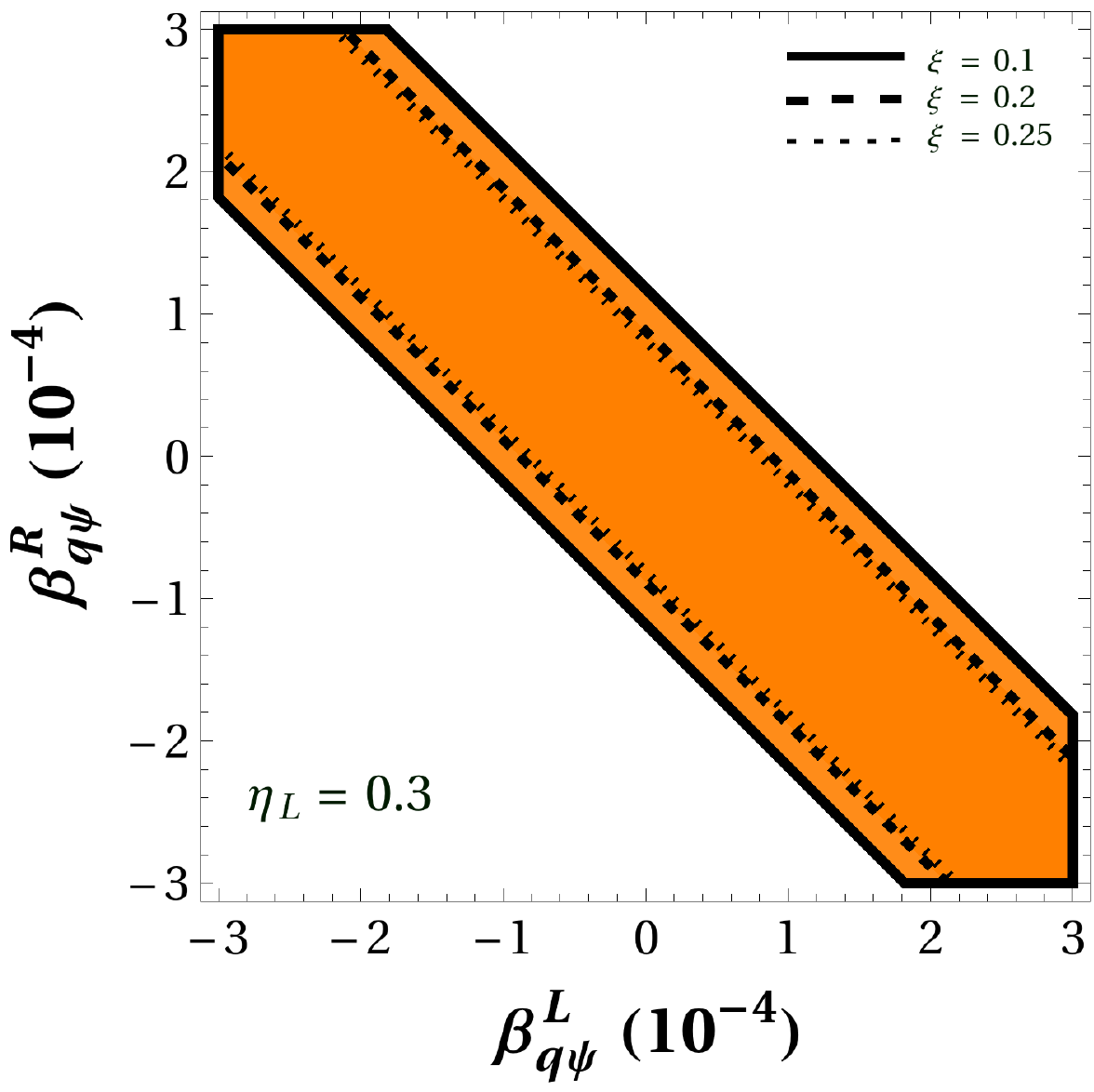}
\includegraphics[scale=0.43]{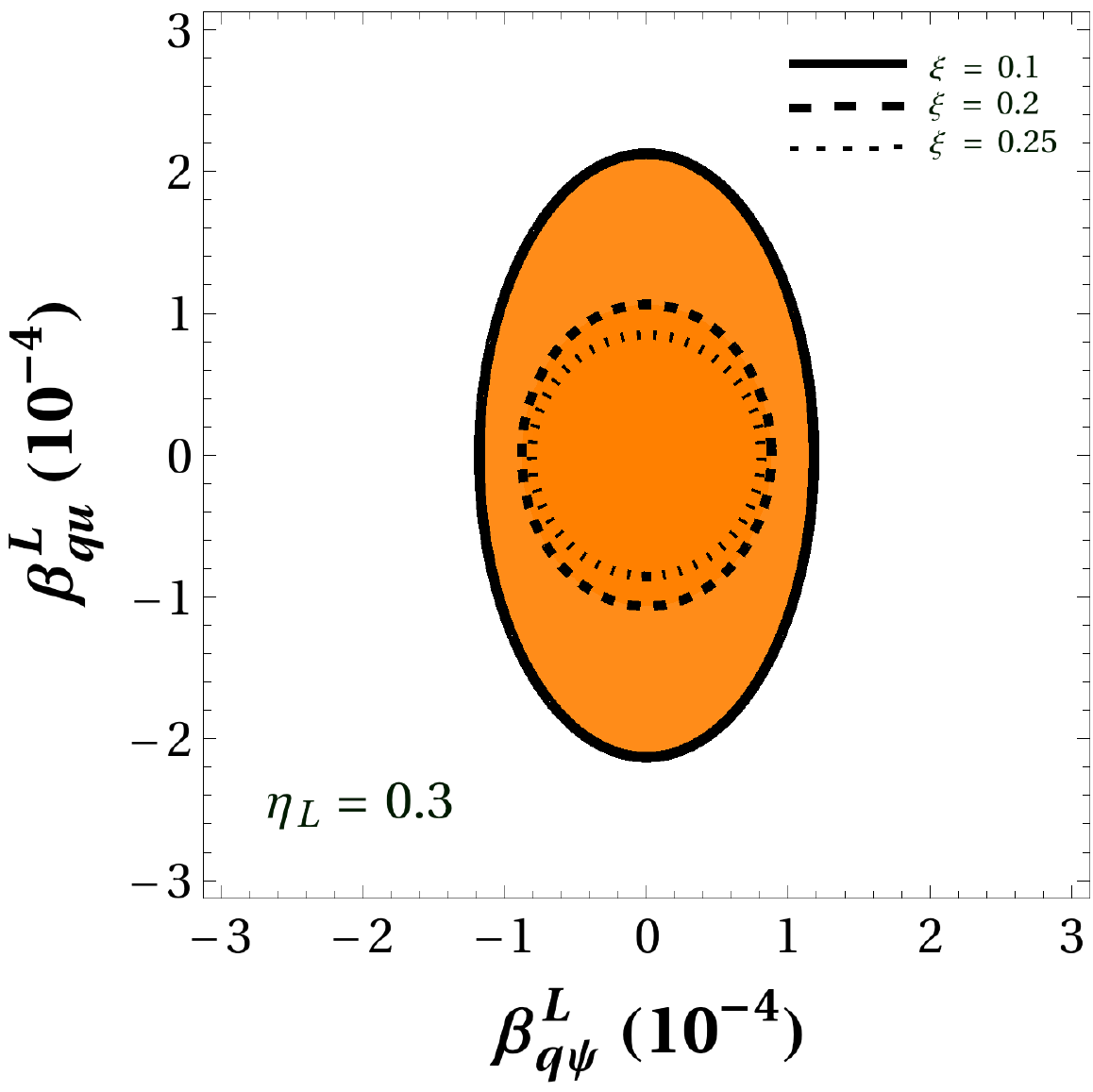}
\caption{\sf Parameter spaces from the experimental neutron EDM limits in~\eqref{Numerical-edm-bound-fA-fB-oA} for $\fA$ (top), $\fB$ (bottom-left) and $\oA$ (bottom-centre and right). The parameters are fixed at $\eta _L=\eta _R=\tilde{\eta }_L=\tilde{\eta }_R=0.3$, while the scan over $\xi=\{0.1,\,0.2,\,0.25\}$ has been performed. See the text for details.}
\label{Effective-edm-bounds}
\end{center}
\end{figure}

\nt Fig.~\ref{Effective-edm-bounds} displays some parameter spaces after scanning along $\xi=\{0.1,\,0.2,\,0.25\}$ through the previous inequalities and simultaneously holding two coefficients on per scan. Only three parameter spaces out of the fifteen possibles in the $\eta_L$-small limit for $\fA$ have been gathered in Fig.~\ref{Effective-edm-bounds} (top), whereas the single one for $\fB$ and two out of the six available for $\oA$ have been displayed for briefness reasons in the bottom-left and bottom-center and right respectively. Strongly constrained coefficients of the order $\cO(10^{-4})$ are allowed, with large opposite-sign contributions cancelling out among them and satisfying bounds in~\eqref{Numerical-edm-bound-fA-fB-oA}. Peculiar situation occurs for $\oA$, where the parameter space is precisely constrained due to the imaginary component of the associated EDM-coefficient in~\eqref{EDM-fA-fB-oA}-\eqref{Numerical-edm-bound-fA-fB-oA}.

\begin{table}[htb!]
\centering
\small{
\hspace*{-1mm}
\renewcommand{\arraystretch}{1.0}
\begin{tabular}{c||c||c||c}
\hline\hline
\\[-4mm]
\bf Model & $\mathbf{\beta _i= \pm 1}$ 
& 
$\begin{array}{l}
\mathbf{\beta_i=\dots= \beta_l= \pm 1}\\[1.5mm]
\mathbf{\beta_m=\beta_n  = \mp 1}   
\end{array}$
& 
\bf Other combinations\\[0.5mm]
\hline\hline
\\[-4mm]   
$\begin{array}{l}
\\[1.5mm]
\fA   
\end{array}$ &  
$\begin{array}{l}
\\[.1cm]
-
\\[-1.5cm]
\end{array}$
& 
$|\eta _R|\approx 0.30$
&
$\eta _R\approx 0.70$
\\[1mm] 
\cline{3-4}\\[-4mm] 
&& $\begin{array}{l}
\\[1mm]
\eta _R\approx 1.0  
\end{array}$ & $\eta _R\approx 1.30$
\\[1mm] 
\cline{4-4}\\[-4mm] 
&&  & $|\eta_R|\lesssim 1.1\times 10^{-4}$
\\[1mm] 
\hline\hline\\[-4mm] 
$\begin{array}{l}
\fB  
\end{array}$ &  
$-$
&  
$-$
&
$|\eta_R|\lesssim 1.1\times 10^{-4}$
\\[1mm] 
\hline\hline\\[-4mm] 
$\oA$
&  
$\tilde{\eta}_R\approx -1.77\times 10^{-5}$
&  
$\tilde{\eta}_R\approx -1.77\times 10^{-5}$
&
$|\tilde{\eta}_R|\approx 0.16$
\\[1mm] 
\hline \hline
\end{tabular}
\caption{\sf Allowed $\eta_R$ and $\tilde{\eta}_R$-ranges from bounds in~\eqref{Bounds-FCNC} for $\xi=0.1$, by setting the $\beta$-coefficients at different values: $\pm 1$ all at once, $\pm 1$ by pairs while the rest at $\mp 1$, and several other combinations.}  
\label{eta-ranges-extra}
}
\end{table}

Calling for the experimental bound in~\eqref{eq:eedm}, the parameters $\eta_R$, $\tilde{\eta}_R$ (and therefore $\tilde{\eta}_L$ assuming $\tilde{\eta}_L \approx \tilde{\eta}_R$) can be restricted via the EDM coefficients obtained in~\eqref{EDM-fA-fB-oA}. Table~\ref{eta-ranges-extra} reports the allowed ranges for $\xi=0.1$ and for different coefficient settings. Models $\fB$ and $\oA$ are disfavoured for any $\beta$-coefficients setting. Conversely, model $\fA$ favours a couple of tensors interfering oppositely with respect to the rest of tensors (3rd column). In fact, at $\fA$ favours the minimal $\eta _R\approx 0.30$ and the maximal one $\eta _R\approx 1.0$ in~\eqref{eta-theoretical-ranges}, while achieving intermediate values of $\eta _R\approx 0.70$ (4th column) or slightly exceeding the predicted one when $\eta _R\approx 1.30$ for the same model. At $\fA$ one can thus infer the relation $y_L=y_R\approx 0.7\,g_{\bf{4}}$  aside from those already pointed out in Table~\ref{eta-ranges}.

\section{Comments on $\rho$ and $\Psi$-production}
\label{Production}

\nt The mass scale $m_\rho$ of a new strongly-interacting sector can be set via searches for direct production of resonances at the LHC. The dominant mechanis is via Drell--Yan processes (see for example Ref.~\cite{Falkowski:2011ua}). A class of theories motivated both theoretically and experimentally is one in which the spin-1 resonance couples to light fermions only through its mixing to the SM gauge fields~\cite{Contino:2006nn} (see Ref.~\cite{Redi:2011zi} for alternative possibilities). In this case the Drell--Yan production cross section scales as $1/g_{\rho}^{2}$, since couplings of the resonances to the SM fermions are suppressed by $1/g_{\rho}$. The strongest exclusion limits are currently set by the LHC searches performed at $8$~TeV with $20$ fb$^{-1}$ in final states with one lepton and missing 
transverse energy~\cite{Chatrchyan:2012rva} or dileptons~\cite{CMS:2013qca,ATLAS:2013jma}, looking for charged and neutral spin-1 resonances respectively.  For values of $\xi$ of order 1, searches for resonances decaying into $WZ$, in particular those with three leptons in the final state~\cite{CMS:2013vda,ATLAS:2013lma}, give slightly stronger bounds.~\footnote{We find that the more recent searches for spin-1 resonances decaying to pairs of vector bosons with boosted decay products~\cite{CMS:2013xea,CMS:2013dfa,CMS:2013fea} give less strong constraints.} Assuming the $\rho$ to be a $({\bf 3},{\bf 1})$ of $SU(2)_{L} \times SU(2)_R$, we translated the bounds on 
$(\sigma \times BR)$ set by the experimental collaborations into a combined exclusion region in the $(\xi, m_{\rho})$ plane. 

Concerning the vector resonance production, the role of spin-0 and spin-1 resonances on the PNGBs scattering were studied in~\cite{Contino:2011np}. Their experimental searches~\cite{ATLAS:2013jma} were explored for $\xi=0.1$ in~\cite{Contino:2013gna,Pappadopulo:2014qza,Greco:2014aza}. Recently, the impact of heavy triplet resonances at the LHC in the final states $l^+l^-$ and $l\nu_l$ ($l=e,\mu$), $\tau^+\tau^-$, $jj$, $t\bar{t}$ as well as on the gauge and gauge-Higgs channels $WZ$, $WW$, $WH$ and $ZH$, has been analysed (see~\cite{CarcamoHernandez:2017pei,Shu:2016exh,Shu:2015cxm} and references therein), constraining the vector resonance mass in the range $2.1-3$ TeV. 

On the other hand, the most stringent experimental constraints on $\fourplet$ and $\singlet$ from the direct searches had been derived in~\cite{Chatrchyan:2013wfa,Chatrchyan:2013uxa}. In fact, by means of pair production mechanism driven mostly by QCD interactions, rough limits on $m_{\Xft}$ and $m_{\Xtt}$ were respectively established as $800$~GeV and $700$~GeV. Experimental searches for the singly produced partners~\cite{ATLAS:2014pga} and searches for pair production into the bounds on singly produced partners~\cite{DeSimone:2012fs,Azatov:2013hya,Matsedonskyi:2014mna,Matsedonskyi:2015dns} have been considered. Additionally, the nineplet case has been analysed yielding $m_9 \gtrsim 1$~TeV~\cite{Matsedonskyi:2014lla}.  

These bounds have been updated and refined following the latest ATLAS and CMS results~\cite{Aaboud:2017qpr,Sirunyan:2017pks} .  The search for the pair production of vector-like top quarks in final states with exactly one lepton, at least four jets and high missing transverse momentum has allowed to exclude masses below 870~GeV (890~GeV expected) and 1.05~TeV (1.06~TeV expected), for the singlet and doublet models respectively. The search was based on $36.1\,\text{fb}^{-1}$ of $\sqrt{s}=13$~TeV LHC $pp$ collision data recorded by ATLAS in 2015 and 2016 (see~\cite{Aaboud:2017qpr} for more details).  CMS has recently analysed double production of vector-like quarks  electrically charged as $2/3$ and $-4/3$, in $pp$ interactions at $\sqrt{s} = 13$~TeV and decaying exclusively via the $bW$ channel.  By comparing these limits with the predicted theoretical cross section of the pair production, the production of the $2/3$ and $-4/3$ electrically charged quarks is excluded
at 95\% confidence level for masses below 1295~GeV (1275~GeV expected).
More generally, the results set upper limits on the product of the
production cross section and branching fraction to $bW$ for any new heavy
quark decaying to this channel (see~\cite{Sirunyan:2017pks}).

A complete study on the mechanisms and production channels for the vector resonances and top partners in our framework is far from the scopes of the present work. It is postponed for a future analysis where the consequences and effects of the interplay $\rho$-$\Psi$ will be accounted. The flexibility entailed by the freedom of our set-up will be useful in dealing with different possible final states concerning the charged and neutral physical vector resonances, and all the physical states brought by the top partners scale as well. 

\section{Summary}
\label{Summary}

\nt In this work we have thoroughly explored the interplay among three matter sectors: elementary, top partners and vector resonances in a $SO(5)$ composite Higgs Model. Such interplay has been explicitly and completely parametrised through interactions of the vector resonance $\rho$, here assumed to be spin-1 triplets of $SU(2)_L\times SU(2)_R$, coupled to the $SO(5)$-invariant fermionic currents and 2nd rank tensors listed in Table~\ref{Fermion-currents-set}, and provided by the first time in this work. These invariants cover all the structures built upon the SM elementary sector and the top partners permitted by the unbroken $SO(4)$, here restricted to the fourplet $\fourplet$ and singlet $\singlet$ embeddings. Such matter content spans four models in~\eqref{Models}, coupled each of them to the $\rho$-resonance and subsequently scanned along $\xi=\{0.1,\,0.2,\,0.25\}$.

The assumption of top partners and vector resonances demands two NP mass scales $M_{\bf{4(1)}}$ and $m_\rho$, below the cut-off of the theory. We consider the ansatz $M_{\bf{4(1)}} < m_{\rho}$ to tackle the low energy effects sourced by our interplaying set-up. Implementing $\rho$-EOM a priori and $\Psi$-EOM a posteriori, we are led to obtain 4-fermion operators whose phenomenological impact has been considered via flavour-dijet processes and EDM bounds. Parametric spaces implied by the coefficients weighting the current and tensors were explored, consequently allowing a rough $\cO(10^{-3}$--$10^{-4})$ and $\cO(10^{-4})$ orders of magnitude respectively. Such parameter spaces allow us to get an insight about the strength of the coefficients that weight our interplaying structures, providing thus an accurate estimation for the EFT approach of top partner-vector resonances undertaken here. The strength of the Goldstone symmetry breaking and the extra couplings brought by the top partner mass scales are linked through the parameters $\eta$, here also estimated accordingly with theoretical expectations. As an important remark, the scenario of opposite interfering currents and tensor turns out to be the most favoured at models $\fA$ and $\fB$, as $\oA$ and $\oB$ lack of the parametric freedom involved by the latter. Production channels for the top partners and vector resonances were briefly commented and left for a soon future analysis accounting for their interplay.

The tantalizing scenario of weak scale naturalness motivated us to model, parametrise and explore the possible interplay that may occur among the elementary, top partners, and vector resonances sectors, together with its low energy implications. These result should be useful when dealing with EFT approaches beyond SM frameworks, specifically in coping with new interaction that might underlie the existence of exotic matter content in our nature, and potentially discovered at future high energy colliders.

\section*{Acknowledgements}

\nt Thanks to J. Shu  for his valuable feedback at the beginning stage of this draft. J.Y. acknowledges the support of Fondecyt (Chile) grant No. 3170480. The work of A.Z. is supported by Conicyt (Chile) grants ACT1406 and PIA/Basal FB0821, and by  Fondecyt (Chile) grant 1160423.

\appendix
\small

\section{CCWZ formalism}
\label{CCWZ}

\nt The $SO(4) \simeq SU(2)_L \times SU(2)_R$ unbroken generators and the broken ones parametrizing the coset $\textrm{SO}(5)/SO(4)$ in the fundamental representation are
\be
(T^a_\chi)_{IJ} = -\frac{i}{2}\left[\frac{1}{2}\varepsilon^{abc}
\left(\delta_I^b \delta_J^c - \delta_J^b \delta_I^c\right) \pm
\left(\delta_I^a \delta_J^4 - \delta_J^a \delta_I^4\right)\right],\qquad
T^{i}_{IJ} = -\frac{i}{\sqrt{2}}\left(\delta_I^{i} \delta_J^5 - \delta_J^{i} \delta_I^5\right)\,,
\label{eq:SO4_gen-SO5/SO4_gen}
\ee

\nt with $\chi= L,\,R$ and $a= 1,2,3$, while $i = 1, \ldots, 4$. The normalization of $T^{A}$'s is chosen as ${\rm Tr}[T^A, T^B] = \delta^{AB}$. The $4\times 4$ matrices $\tau^a$ appearing in the bilinear fourplets at $\fA$ and $\fB$ in Table~\ref{Fermion-currents-set} are defined as
\be
\left[T^{a},T^{i}\right]=\left(t^{a}\right)_{{j}{i}}T^{j}\,.
\label{taus}
\ee

\nt Gauging the SM subgroup of $SO(5)$ requires us to introduce local transformations via $U$ matrices that will couple the SM gauge fields to the composite resonances. The CCWZ $d$ and $e$ symbols are in order to do so 
\be
-U^t [A_\mu + i \partial_\mu ] U = d_\mu^{\hat a} T^{\hat a} + e^{a}_{\mu} T^a + e^{X}_{\mu}
\label{d-e}
\ee

\nt where $A_{\mu}$ stands for ${\cal G}_{\text{SM}}$ gauge fields 
\be
\begin{aligned}
A_\mu &= \frac{g}{\sqrt{2}}W^+_\mu T^-_L +\frac{g}{\sqrt{2}}W^-_\mu T^+_L+ g \left(\cw Z_\mu+\sw A_\mu \right)T_L^3+g' \left(\cw A_\mu-\sw Z_\mu \right)(T_R^3+Q_X)
\label{gfd}
\end{aligned}
\ee

\nt with $T^\pm_\chi=\left(T^1_\chi\mp i T^2_\chi\right)/\sqrt{2}$, the implied notation $(\cw,\,\sw)\equiv(\cos\theta_{\rm w},\,\sin\theta_{\rm w})$, and the SM couplings of $SU(2)_L$ and $\textrm{U}(1)_Y$, $g$, $g'$ respectively, where $Q_X$ is the $X$-charge matrix. The definition~\eqref{d-e} can be expanded in fields as
\be
d_\mu^i=\frac{\sqrt{2}}{f}(D_\mu h)^i+{\cal O} (h^3),\qquad
e^{a}_\mu = -A^{a}_\mu-\frac{i}{f^2}(h{\lra{D}_\mu}h)^a+{\cal O} (h^4),\qquad
e^{X}_\mu = -g^{\prime}  Q_X B_\mu \,,
\label{CCWZ-symbols}
\ee

\nt with $B_{\mu}$ the $U(1)_Y$ gauge boson. Covariant derivatives acting on the composite sector fields are built out of $e$ symbols. For the $\Psi$ field transforming in the fundamental representation of $SO(4)$ one has
\be
\nabla_\mu\Psi \,=\,D_\mu\Psi+i\,e_{\mu}^at^a\Psi\,.
\label{covariant-derivative}
\ee
The term $\slashed{e}=e_\mu\gamma^\mu$ is included in $\LL_{\text{comp}}$ to fully guarantee the $SO(5)$ invariance. Strength field tensors are straightforwardly introduced as
\be
e_{\mu \nu} = \partial_{\mu} e_{\nu} - \partial_{\nu} e_{\mu} + i g_{\rho} [e_{\mu},e_{\nu}], \qquad\qquad
e_{\mu \nu}^{X} = \partial_{\mu} e^X_{\nu} - \partial_{\nu} e^X_{\mu}.
\ee

\nt Finally, the covariant derivatives $D_\mu$ associated to each one of the elementary fields as well to the corresponding top partner are given by

\bea
D_\mu\,q_L&=&\left(\partial_\mu-ig W_\mu^i {\sigma^i\over 2}-i{1\over 6}g' B_\mu-i\,g_SG_\mu\right)q_L\,, \\
D_\mu\,u_R&=&\left(\partial_\mu-i{2\over 3}g' B_\mu-i\,g_SG_\mu\right)u_R \,,\\
D_\mu\fourplet &=& \left(\partial_\mu-i {2\over 3} g' B_\mu -i\,g_SG_\mu\right)\fourplet\,.
\label{cder}
\eea

\nt with $g,g'$ and $g_S$ the ${\textrm{SU}}(2)_L\times {\textrm{U}}(1)_Y$ and ${\textrm{SU}}(3)_c$ gauge couplings. Notice the gluon presence in the last covariant derivative as the top 
partners form a color triplet.

\section{Fermion currents in $SU(2)_L\times SU(2)_R$-components}
\label{Fermion currents}

\nt All the fermion currents listed in Table~\ref{Fermion-currents-set} are provided here after the vacuum alignment and before using the rotations that diagonalize the mass-matrices. In the $SU(2)_L\times SU(2)_R$-components, we have for the model \fA 
\small{
\begin{align}
\label{fA-Jq-L}
& \cJmuuqL =\,\frac{1}{4}\left(c^+_{\theta } \cJ^\mu_{t_Lb_L},\,\,-i\,c^+_{\theta }\,\cJ^\mu_{t_Lb_L},\,\,c_{\theta }\,\cJ^\mu_{t_Lt_L}-\cJ^\mu_{b_Lb_L}\right)\,+\,\hc\,,\\ \nn \\
\label{fA-Jq-R}
& \cJmuuqR =\,-\frac{1}{4}\left(c^-_{\theta } \cJ^\mu_{t_Lb_L},\,\,-i\,c^-_{\theta }\,\cJ^\mu_{t_Lb_L},\,\,c_{\theta }\, \cJ^\mu_{t_Lt_L}+\cJ^\mu_{b_Lb_L}\right)\,+\,\hc\,,
\end{align}
}

\small{
\begin{align}
\label{fA-Jpsi-L}
& \cJmuupsiL =\frac{1}{2}\left(\cJ^\mu_{\mathcal{T}\mathcal{B}}+
\cJ^\mu_{X_{\text{5/3}}X_{\text{2/3}}},-i \left(\cJ^\mu_{\mathcal{T}\mathcal{B}}+\cJ^\mu_{X_{\text{5/3}}X_{\text{2/3}}}
\right),\frac{1}{2}\left(\cJ^\mu_{\mathcal{T}\mathcal{T}}
-\cJ^\mu_{\mathcal{B}\mathcal{B}}+\cJ^\mu_{X_{\text{5/3}}X_{\text{5/3}}}
-\cJ^\mu_{X_{\text{2/3}}X_{\text{2/3}}}\right)\right)+\hc,\\ \nn \\
\label{fA-Jpsi-R}
& \cJmuupsiR =\frac{1}{2}\left(\cJ^\mu_{X_{\text{2/3}}\mathcal{B}}+
\cJ^\mu_{X_{\text{5/3}}\mathcal{T}},-i \left(\cJ^\mu_{X_{\text{2/3}}\mathcal{B}}
+\cJ^\mu_{X_{\text{5/3}}\mathcal{T}}\right),\frac{1}{2}\left(-\cJ^\mu_{\mathcal{T}\mathcal{T}}-
\cJ^\mu_{\mathcal{B}\mathcal{B}}+
\cJ^\mu_{X_{\text{5/3}}X_{\text{5/3}}}+
\cJ^\mu_{X_{\text{2/3}}X_{\text{2/3}}}\right)\right)+\hc,\\ \nn \\
\label{fA-Jupsi-L-R}
& \cJmuuupsiL =-\cJmuuupsiR =\,\frac{s_{\theta }}{2 \sqrt{2}}\left(\cJ^\mu_{t_R\mathcal{B}_R}-
\cJ^\mu_{X_{\text{5/3}R}t_R},\,\,-i \left(\cJ^\mu_{t_R\mathcal{B}_R}
-\cJ^\mu_{X_{\text{5/3}R}t_R}\right),\,\,\cJ^\mu_{\mathcal{T}_Rt_R}+\cJ^\mu_{X_{\text{2/3}R}t_R}\right)\,+\,\hc\,,\\ \nn \\ 
\label{fA-Jqpsi-L}
& \cJmuuqpsiL =\,\frac{1}{4}\left(c^+_{\theta } \cJ^\mu_{t_L\mathcal{B}_L}-c^-_{\theta } \cJ^\mu_{X_{\text{5/3}L}t_L}+2 \cJ^\mu_{\mathcal{T}_Lb_L},\,\,-i\left(c^+_{\theta } \cJ^\mu_{t_L\mathcal{B}_L}-c^-_{\theta } \cJ^\mu_{X_{\text{5/3}L}t_L}+2 \cJ^\mu_{\mathcal{T}_Lb_L}\right),\right. \nn \\ \nn \\
& \phantom{ =\,\frac{1}{4}c^+_{\theta }\cJ^\mu_{t_L\mathcal{B}_L}}\left. c^+_{\theta } \cJ^\mu_{\mathcal{T}_Lt_L}+c^-_{\theta } \cJ^\mu_{X_{\text{2/3}L}t_L}-2 \cJ^\mu_{\mathcal{B}_Lb_L}\right)\,+\,\hc\,,
\end{align}
}

\small{
\begin{align}
\label{fA-Jqpsi-R}
& \cJmuuqpsiR =\,\frac{1}{4}\left(-c^-_{\theta } \cJ^\mu_{t_L\mathcal{B}_L}+c^+_{\theta } \cJ^\mu_{X_{\text{5/3}L}t_L}+2 \cJ^\mu_{X_{\text{2/3}L}b_L},\,\,i\left(c^-_{\theta } \cJ^\mu_{t_L\mathcal{B}_L}-c^+_{\theta } \cJ^\mu_{X_{\text{5/3}L}t_L}-2 \cJ^\mu_{X_{\text{2/3}L}b_L}\right),\right. \nn \\ \nn \\
& \phantom{ =\,\frac{1}{4}c^+_{\theta }\cJ^\mu_{t_L\mathcal{B}_L}}\left.-\left(c^+_{\theta } \cJ^\mu_{\mathcal{T}_Lt_L}+c^-_{\theta } \cJ^\mu_{X_{\text{2/3}L}t_L}+2 \cJ^\mu_{\mathcal{B}_Lb_L}\right)\right)\,+\,\hc\,,
\end{align}
}

\nt where each one of the currents above $\cJ^\mu_{\psi\phi}$ are defined as
\be
\cJ^\mu_{\psi_\chi\phi_\chi}= \bar\psi_\chi \gamma^\mu\phi_\chi\qquad\qquad 
\chi= L,\,R
\label{Generic-currents}
\ee

\nt with the coefficients
\be
c^\pm_{n\theta }\equiv c_{n\theta } \pm 1,\qquad\qquad c_{n\theta }\equiv \cos(n\,\theta)
\label{Simplified-coefficients}
\ee

\nt For the case of \oA, the currents $\cJmuuq$ are the same as the first two in~\eqref{fA-Jq-L}-\eqref{fA-Jq-R}, whilst $\cJmuuqpsi$ are
\be
\begin{aligned}
& \cJmuuqpsiL \equiv \cJmuuqpsiR =\,\frac{1}{\sqrt{2}}\left(i\,\cJ^\mu_{\tilde{\mathcal{T}}_Lb_L},\,\, \cJ^\mu_{\tilde{\mathcal{T}}_Lb_L},\,\,i\,\cJ^\mu_{\tilde{\mathcal{T}}_Lt_L}\right)\,+\,\hc\,.
\end{aligned}
\label{Currents-oA-components}
\ee

\nt For the model $\fB$ we have
\small{
\begin{align}
\label{fB-Jq-L}
& \cJmuuqL =\frac{1}{4}\left(\frac{1}{2} \left(c_{\theta }+c_{2 \theta }+c_{3 \theta }+1\right) \cJ^\mu_{t_Lb_L},-\frac{i}{2} \left(c_{\theta }+c_{2 \theta }+c_{3 \theta }+1\right) \cJ^\mu_{t_Lb_L},\,c_{\theta } \left(c_{2 \theta } \cJ^\mu_{t_Lt_L}-c_{\theta } \cJ^\mu_{b_Lb_L}\right)\right)+\hc,
\end{align}
}

\small{
\begin{align}
\label{fB-Jq-R} 
& \cJmuuqR =\,\frac{1}{2}\left(\left(c_{\theta }+c_{2 \theta }+1\right) s_{\frac{\theta }{2}}^2 \cJ^\mu_{t_Lb_L},\,\,-i \left(c_{\theta }+c_{2 \theta }+1\right) s_{\frac{\theta }{2}}^2 \cJ^\mu_{t_Lb_L},\,\,-\frac{c_{\theta }}{2} \left(c_{\theta } \cJ^\mu_{b_Lb_L}+c_{2 \theta } \cJ^\mu_{t_Lt_L}\right)\right)\,+\,\hc\,,\\ \nn \\
\label{fB-Jqpsi-L} 
& \cJmuuqpsiL =\,\frac{1}{4}\left(i \left(c_{2 \theta } \left(\cJ^\mu_{t_L\mathcal{B}_L}+\cJ^\mu_{X_{\text{5/3}L}t_L}\right)
+c_{\theta } \left(-2\cJ^\mu_{\mathcal{T}_Lb_L}+\cJ^\mu_{t_L\mathcal{B}_L}
-\cJ^\mu_{X_{\text{5/3}L}t_L}\right)\right),\right.\nn \\ \nn \\
& \phantom{ =\,\frac{1}{4}c^+_{\theta }\cJ^\mu_{t_L\mathcal{B}_L}}\left. c_{2 \theta } \left(\cJ^\mu_{t_L\mathcal{B}_L}+\cJ^\mu_{X_{\text{5/3}L}t_L}\right)
+c_{\theta } \left(-2 \cJ^\mu_{\mathcal{T}_Lb_L}+
\cJ^\mu_{t_L\mathcal{B}_L}-\cJ^\mu_{X_{\text{5/3}L}t_L}\right),\right.\nn \\ \nn \\
& \phantom{ =\,\frac{1}{4}c^+_{\theta }\cJ^\mu_{t_L\mathcal{B}_L}}\left. i \left(c_{\theta } \left(-\cJ^\mu_{\mathcal{T}_Lt_L}+\cJ^\mu_{X_{\text{2/3}L}t_L}+2 \cJ^\mu_{\mathcal{B}_Lb_L}\right)-c_{2 \theta } \left(\cJ^\mu_{\mathcal{T}_Lt_L}
+\cJ^\mu_{X_{\text{2/3}L}t_L}\right)\right)\right)\,+\,\hc\,,\\ \nn \\
\label{fB-Jqpsi-R} 
& \cJmuuqpsiR =\,\frac{1}{4}\left(i \left(c_{\theta } \left(\cJ^\mu_{t_L\mathcal{B}_L}-2 \cJ^\mu_{X_{\text{2/3}L}b_L}-\cJ^\mu_{X_{\text{5/3}L}t_L}\right)-c_{2 \theta } \left(\cJ^\mu_{t_L\mathcal{B}_L}+\cJ^\mu_{X_{\text{5/3}L}t_L}\right)
\right),\right.\nn\\ \nn \\
& \phantom{ =\,\frac{1}{4}c^+_{\theta }\cJ^\mu_{t_L\mathcal{B}_L}}\left. c_{\theta } \left(\cJ^\mu_{t_L\mathcal{B}_L}-2 \cJ^\mu_{X_{\text{2/3}L}b_L}-\cJ^\mu_{X_{\text{5/3}L}t_L}\right)-c_{2 \theta } \left(\cJ^\mu_{t_L\mathcal{B}_L}+\cJ^\mu_{X_{\text{5/3}L}t_L}\right)
,\right.\nn\\ \nn \\
& \phantom{ =\,\frac{1}{4}c^+_{\theta }\cJ^\mu_{t_L\mathcal{B}_L}}\left.-i \left(c_{\theta } \left(-\cJ^\mu_{\mathcal{T}_Lt_L}+\cJ^\mu_{X_{\text{2/3}L}t_L}-2 \cJ^\mu_{\mathcal{B}_Lb_L}\right)-c_{2 \theta } \left(\cJ^\mu_{\mathcal{T}_Lt_L}
+\cJ^\mu_{X_{\text{2/3}L}t_L}\right)\right)\right)\,+\,\hc
\end{align}
}

\nt where the currents $\cJmuupsi$ are already listed as the third and fourth one in~\eqref{fA-Jpsi-L}-\eqref{fA-Jpsi-R}. Finally, the single currents for the model \oB, \eg 
$\cJmuuq$, correspond to the first two in~\eqref{fB-Jq-L}-\eqref{fB-Jq-R}.

\section{$\eta$-large limit and $\cO_{u_L\,u_R}$\,\,\,\&\,\,\,$\cO_{u_R\,d_L}$}
\label{Extra-4f-operators}

\nt Higher order terms become relevant when the parameters $\eta_{L(R)}$ and $\tilde{\eta}_{L(R)}$ are large. This case has rather long expressions for the Wilson coefficients in Table~\ref{Wilson-coefficients}. The parameter spaces become wider as more operator coefficients play a role in them. The 4-fermion operators in~\eqref{4-fermion-ops-extra} emerge at $\cO(\eta^3)$ and $\cO(\eta^4)$-order for $\fA$ and $\fB$ respectively, having the corresponding Wilson coefficients
\be
\begin{aligned}
&c_{u_L\,u_R} = \frac{\xi}{4}\,\eta _L \left(\eta _L\,\eta _R-\frac{M_{\bf{1}}}{M_{\bf{4}}}\,\tilde{\eta }_L\,\tilde{\eta }_R\right) \left(\alpha^L_q\,\alpha^L_{u \psi}+\alpha^R_q\, \alpha^R_{u \psi}\right),\\
&c_{u_R\,d_L} = -\frac{\xi}{4}\,\eta _L \left(\eta _L\,\eta _R-\frac{M_{\bf{1}}}{M_{\bf{4}}}\,\tilde{\eta }_L\,\tilde{\eta }_R\right) \left(\alpha^L_q\,\alpha^L_{u \psi}-\alpha^R_q\,\alpha^R_{u \psi}\right),
\label{Extra-Wilson-coeffcients-fA}
\end{aligned}
\ee

\nt at $\fA$, and 

\be
\begin{aligned}
&c_{u_L\,u_R} = \frac{\xi}{2}\,\eta _L^2\left(\frac{M_{\bf{1}}}{M_{\bf{4}}}\,\tilde{\eta }_R+\eta _R\right)^2\left[\alpha^L_\psi\,\alpha^L_q  +  \alpha^R_q\,\alpha^R_\psi\right],\\
&c_{u_R\,d_L} = -\frac{\xi}{2}\,\eta _L^2 \left(\frac{M_{\bf{1}}}{M_{\bf{4}}}\,\tilde{\eta }_R + \eta _R\right)^2 \left(\alpha^L_\psi\,\alpha^L_q-\alpha^R_q\,\alpha^R_\psi\right)
\label{Extra-Wilson-coeffcients-fB}
\end{aligned}
\ee

\nt at $\fB$. By imposing the bounds in~\cite{Calibbi:2012at} on the operator $\cO_{u_L\,u_R}$, and  additionally setting  $\eta _L=\eta _R=\tilde{\eta }_L=\tilde{\eta }_R=0.3$, together with $M_{\bf{4}}\sim 2 M_{\bf{1}}$, a rough order of magnitude of $10^{-1}$ is obtained for the coefficients $\alpha$'s, entailed by the extra suppression from the higher $\eta$-powers.
Table~\ref{eta-ranges-dijet-extra} reports allowed $\eta$-ranges by assuming the scenario $\eta _L=\eta _R$ and $\tilde{\eta }_L=\tilde{\eta }_R$, with $\eta_L=0.3$ following the minimal value in~\eqref{eta-theoretical-ranges}. Different coefficient's settings as the ones provided yield a constant value for the corresponding Wilson coefficients.
\begin{table}
\centering
\small{
\hspace*{-1mm}
\renewcommand{\arraystretch}{1.0}
\begin{tabular}{c||c||c}
\hline\hline
\\[-4mm]
\bf Model & $\alpha _i= \pm 1$ & $\alpha_i=\alpha_j = \pm 1,\, \alpha_k = \alpha_l = \mp 1$
\\[0.5mm]
\hline\hline
\\[-4mm]   
$\begin{array}{l}
\fA   
\end{array}$ &  
$\begin{array}{l}
\\[-2.5mm] 
|\tilde{\eta }_L|=|\tilde{\eta }_R|\approx 0.42
\end{array}$
&  
$\begin{array}{l}
\\[-2.5mm] 
|\tilde{\eta }_L|=|\tilde{\eta }_R|\approx 0.42
\end{array}$\\[4mm] 
\hline\\[-4mm] 
$\begin{array}{l}
\\[-3mm] 
\fB
\end{array}$ &  
$\begin{array}{l}
\\[-2.5mm] 
-0.61\leq \tilde{\eta }_L,\,\tilde{\eta }_R\leq -0.58
\end{array}$
& 
$\begin{array}{l}
\\[-2.5mm] 
-0.61\leq \tilde{\eta }_L,\,\tilde{\eta }_R\leq -0.58
\end{array}$\\[3mm]
\hline \hline
\end{tabular}
\caption{\sf Allowed $\eta$-ranges from dijet bounds in~\eqref{Coeff-dijet-bounds} and by setting $\alpha$-coefficients at 1 for $\xi=0.1$ only, and assuming the scenario $\eta _L=\eta _R$ and $\tilde{\eta }_L=\tilde{\eta }_R$, by fixing $\eta_L=0.3$ accordingly with the minimal inferred value in~\eqref{eta-theoretical-ranges}.} 
\label{eta-ranges-dijet-extra}
}
\end{table}

\section{Top partners EOM}
\label{Partners-EOM}

\nt For a heavy top partners mass scenario the corresponding fields may be integrated them out from the physical spectrum, via their associated EOM. Concerning the $\fA$ and $\oA$ models altogether, and after diagonalizing the mass terms at $\LL_{\text{comp}}+\LL_{\text{mix}}$ in~\eqref{fA-oA-comp}-\eqref{fA-oA-mix} to the physical sector, we can obtain the following field redefinitions for the left handed components as
\beq
\begin{aligned}
&\mathcal{T}_L\to -\frac{\xi}{4}\,\frac{\eta_L}{\eta _L^2+1}\, t_L\,,\quad\quad
&&\quad X_{\text{5/3},L}\to 0\,,\quad\quad 
&&\tilde{\mathcal{T}}_L\to \sqrt{\frac{\xi}{2}}\,\frac{ \tilde{\eta }_L}{\left(\tilde{\eta }_R^2+1\right)\sqrt{\eta _L^2+1}}\,t_L \,,  \\[4mm]
&\quad\mathcal{B}_L\to 0 \,,\quad\qquad
&&X_{\text{2/3},L}\to \frac{\xi}{4}\,\frac{\eta _L}{\sqrt{\eta _L^2+1}}\,t_L\,,\quad\quad 
&&  \\
\end{aligned}
\label{EOM-fA-oA-L}
\eeq

\nt while for the right handed components one obtains
\beq
\begin{aligned}
&\mathcal{T}_R\to \sqrt{\frac{\xi}{2}}\,\frac{\eta _R }{\left(\eta _L^2+1\right) \sqrt{\tilde{\eta }_R^2+1}}\,t_R\,,\quad\quad
&&\quad X_{\text{5/3},R}\to 0\,,\quad\quad 
&&\tilde{\mathcal{T}}_R\to -\frac{\xi}{2}\,\frac{\tilde{\eta }_R}{\tilde{\eta }_R^2+1}\,t_R\,,  \\[4mm]
&\quad\mathcal{B}_R\to -\tilde{\eta }_R\,b_R \,,\quad\quad
&&\quad X_{\text{2/3},R}\to -\sqrt{\frac{\xi}{2}}\,\frac{\eta _R }{\sqrt{\tilde{\eta }_R^2+1}}\,t_R\,.\quad\quad 
&&  \\
\end{aligned}
\label{EOM-fA-oA-R}
\eeq

\nt where the parameters $\eta_{L(R)}$ are defined through

\be
\eta _{L(R)}\to \frac{y_{L(R)} \mathit{f}}{M_{\bf{4}}}\equiv  \frac{y_{L(R)}}{g_{\bf{4}}},\qquad\qquad
\tilde{\eta}_{L(R)}\to \frac{\tilde{y}_{L(R)} \mathit{f}}{M_{\bf{1}}}\equiv  \frac{\tilde{y}_{L(R)}}{g_{\bf{1}}}
\label{eta-parameters}
\ee

\nt and the definition in~\eqref{g-rho-gpartner} has been used. Likewise, from the Lagrangians in~\eqref{fB-oB-comp}-\eqref{fB-oB-mix}, field redefinitions from $\fB$ and $\oB$ for the left handed components as

\beq
\begin{aligned}
&\mathcal{T}_L\to -\frac{5\,\xi}{4}\,\frac{\eta_L}{\eta _L^2+1}\, t_L\,,\quad
\quad X_{\text{5/3},L}\to 0\,,\quad
\tilde{\mathcal{T}}_L\to -\frac{\sqrt{2\,\xi }}{\sqrt{\eta _L^2+1}}\left(\tilde{\eta }_L-\tilde{\eta }_R^2 - \frac{M_{\bf{4}}}{M_{\bf{1}}} \eta _R \tilde{\eta }_R\right)\,t_L\,,  \\[4mm]
&\quad\mathcal{B}_L\to -\frac{\xi}{2}\,\frac{\eta_L}{\eta _L^2+1}\,b_L \,,\quad
\quad X_{\text{2/3},L}\to \frac{3\,\xi}{4}\,\frac{\eta _L}{ \sqrt{\eta _L^2+1}}\,t_L\,,\quad 
&&  \\[4mm] 
\end{aligned}
\label{EOM-fB-oB-L}
\eeq

\nt while for the right handed components one obtains
\beq
\begin{aligned}
&\mathcal{T}_R\to -\sqrt{2\,\xi }\,\,\frac{\sqrt{\tilde{\eta }_R^2+1}}{\eta _L^2+1}\,\eta _L\left(\frac{M_{\bf{1}}}{M_{\bf{4}}}\tilde{\eta }_R + \eta _R \right)\,t_R\,,\quad\quad
&&\quad X_{\text{5/3},L}\to 0\,,\quad\quad 
&&\quad\tilde{\mathcal{T}}_R\to -\tilde{\eta }_R\,t_R\,,  \\[4mm]
&\quad\mathcal{B}_R\to -\tilde{\eta }_R\,b_R\,,\quad\quad
&&\quad X_{\text{2/3},R}\to 0\,.\quad\quad 
&&  \\[4mm] 
\end{aligned}
\label{EOM-fB-oB-R}
\eeq

\nt with the coefficients $\eta$ similarly defined as in~\eqref{eta-parameters}, but with the associated Yukawa couplings $y$ and $\tilde{y}$ corresponding to those for the $\fB$ and $\oB$ models. 

\providecommand{\href}[2]{#2}\begingroup\raggedright\endgroup


\begin{thebibliography}{10}

\bibitem{Aad:2012tfa}
  G.~Aad {\it et al.} [ATLAS Collaboration],
  Phys.\ Lett.\ B {\bf 716} (2012) 1,
  [arXiv:1207.7214 [hep-ex]].
  

  \bibitem{Chatrchyan:2012xdj}
  S.~Chatrchyan {\it et al.} [CMS Collaboration],
  Phys.\ Lett.\ B {\bf 716} (2012) 30,
  [arXiv:1207.7235 [hep-ex]].
  
 
\bibitem{Kaplan:1983fs}
  D.~B.~Kaplan and H.~Georgi,
  Phys.\ Lett.\  {\bf 136B} (1984) 183.  

\bibitem{Kaplan:1983sm}
  D.~B.~Kaplan, H.~Georgi and S.~Dimopoulos,
  Phys.\ Lett.\  {\bf 136B} (1984) 187.
  

\bibitem{Georgi:1984ef}
  H.~Georgi, D.~B.~Kaplan and P.~Galison,
  Phys.\ Lett.\  {\bf 143B} (1984) 152.

\bibitem{Banks:1984gj}
  T.~Banks,
  Nucl.\ Phys.\ B {\bf 243} (1984) 125.

\bibitem{Georgi:1984af}
  H.~Georgi and D.~B.~Kaplan,
  Phys.\ Lett.\  {\bf 145B} (1984) 216.
  
\bibitem{Dugan:1984hq}
  M.~J.~Dugan, H.~Georgi and D.~B.~Kaplan,
  Nucl.\ Phys.\ B {\bf 254} (1985) 299.

  
  \bibitem{Contino:2003ve} 
  R.~Contino, Y.~Nomura and A.~Pomarol,
  Nucl.\ Phys.\ B {\bf 671}, 148 (2003)
  [hep-ph/0306259].
  
  \bibitem{Agashe:2004rs} 
  K.~Agashe, R.~Contino and A.~Pomarol,
  Nucl.\ Phys.\ B {\bf 719}, 165 (2005)
  [hep-ph/0412089].
 

\bibitem{Contino:2011np} 
  R.~Contino, D.~Marzocca, D.~Pappadopulo and R.~Rattazzi,
  JHEP {\bf 1110}, 081 (2011)
  [\hhref{1109.1570}[hep-ph]].


\bibitem{Matsedonskyi:2014iha}
  O.~Matsedonskyi,
  ``On Flavour and Naturalness of Composite Higgs Models'',
  JHEP {\bf 1502} (2015) 154
  [arXiv:1411.4638 [hep-ph]]. 
  
 
\bibitem{Ciuchini:2013pca}
  M.~Ciuchini, E.~Franco, S.~Mishima and L.~Silvestrini,
  JHEP {\bf 1308} (2013) 106
  [arXiv:1306.4644 [hep-ph]].    
  
  \bibitem{Contino:2010rs} 
  R.~Contino,
  [\hhref{1005.4269}[hep-ph]].
  
  
\bibitem{Contino:2006qr} 
  R.~Contino, L.~Da Rold and A.~Pomarol,
  Phys.\ Rev.\ D {\bf 75}, 055014 (2007)
  [hep-ph/0612048].
  
  \bibitem{Matsedonskyi:2012ym} 
  O.~Matsedonskyi, G.~Panico and A.~Wulzer,
  JHEP {\bf 1301}, 164 (2013)
[\hhref{1204.6333}[hep-ph]].
  
  \bibitem{Marzocca:2012zn} 
  D.~Marzocca, M.~Serone and J.~Shu,
  JHEP {\bf 1208}, 013 (2012)
  [\hhref{1205.0770}[hep-ph]].
  
  \bibitem{Pomarol:2012qf} 
  A.~Pomarol and F.~Riva,
  JHEP {\bf 1208}, 135 (2012)
  [\hhref{1205.6434}[hep-ph]].
  
  \bibitem{Redi:2012ha} 
  M.~Redi and A.~Tesi,
  JHEP {\bf 1210}, 166 (2012)
  [\hhref{1205.0232}[hep-ph]].
  
  \bibitem{Panico:2012uw} 
  G.~Panico, M.~Redi, A.~Tesi and A.~Wulzer,
  JHEP {\bf 1303}, 051 (2013)
  [\hhref{1210.7114}[hep-ph]].
  

  \bibitem{Contino:2008hi} 
  R.~Contino and G.~Servant,
  JHEP {\bf 0806}, 026 (2008)
  [\hhref{0801.1679}[hep-ph]].
    
  
\bibitem{Mrazek:2009yu} 
  J.~Mrazek and A.~Wulzer,
  Phys.\ Rev.\ D {\bf 81}, 075006 (2010)
  [\hhref{0909.3977}[hep-ph]].
  
  
\bibitem{DeSimone:2012fs} 
  A.~De Simone, O.~Matsedonskyi, R.~Rattazzi and A.~Wulzer,
  JHEP {\bf 1304}, 004 (2013)
  [\hhref{1211.5663}[hep-ph]].
  
    \bibitem{ewpt}
  C.~Grojean, O.~Matsedonskyi and G.~Panico,
  JHEP {\bf 1310}, 160 (2013)
  [\hhref{1306.4655}[hep-ph]].
  

  \bibitem{Barbieri:2012tu} 
  R.~Barbieri, D.~Buttazzo, F.~Sala, D.~M.~Straub and A.~Tesi,
  JHEP {\bf 1305}, 069 (2013)
  [\hhref{1211.5085}[hep-ph]].
  
  \bibitem{Redi:2012uj} 
  M.~Redi,
  Eur.\ Phys.\ J.\ C {\bf 72}, 2030 (2012)
  [\hhref{1203.4220}[hep-ph]].
  
    \bibitem{KerenZur:2012fr} 
  B.~Keren-Zur, P.~Lodone, M.~Nardecchia, D.~Pappadopulo, R.~Rattazzi and L.~Vecchi,
  Nucl.\ Phys.\ B {\bf 867}, 394 (2013)
  [\hhref{1205.5803}[hep-ph]].


\bibitem{CarcamoHernandez:2017pei}
  A.~E.~Carcamo Hernández, B.~D.~Sáez, C.~O.~Dib and A.~Zerwekh,
  arXiv:1707.05195 [hep-ph].  
  
  
\bibitem{Hernandez:2015xka}
  A.~E.~Carcamo Hernandez, C.~O.~Dib and A.~R.~Zerwekh,
  Nucl.\ Part.\ Phys.\ Proc.\  {\bf 267-269} (2015) 35
  [arXiv:1503.08472 [hep-ph]].

  
 \bibitem{ccwz} 
S. Coleman, J. Wess and B. Zumino, Phys. Rev. 177 (1969) 2239; 
C. Callan, S. Coleman, J. Wess and B. Zumino, 
Phys. Rev. 177 (1969) 2247.


\bibitem{Contino:2013gna}
  R.~Contino, C.~Grojean, D.~Pappadopulo, R.~Rattazzi and A.~Thamm,
  JHEP {\bf 1402} (2014) 006
  [arXiv:1309.7038 [hep-ph]].  


\bibitem{Agashe:2006at}
  K.~Agashe, R.~Contino, L.~Da Rold and A.~Pomarol,
  Phys.\ Lett.\ B {\bf 641} (2006) 62
  [hep-ph/0605341].
  
 \bibitem{Mrazek:2011iu}
  J.~Mrazek, A.~Pomarol, R.~Rattazzi, M.~Redi, J.~Serra and A.~Wulzer,
  Nucl.\ Phys.\ B {\bf 853} (2011) 1
  [\hhref{1105.5403}[hep-ph]].


\bibitem{YZ}
  J.~Yepes and A.~Zerwekh, in progress.    
  

\bibitem{Ecker:1989yg}
  G.~Ecker, J.~Gasser, H.~Leutwyler, A.~Pich and E.~de Rafael, ``Chiral Lagrangians for Massive Spin 1 Fields, Phys.\ Lett.\ B {\bf 223} (1989) 425.



\bibitem{Bando:1984ej}
  M.~Bando, T.~Kugo, S.~Uehara, K.~Yamawaki and T.~Yanagida,
  Phys.\ Rev.\ Lett.\  {\bf 54} (1985) 1215.

\bibitem{Casalbuoni:1985kq}
  R.~Casalbuoni, S.~De Curtis, D.~Dominici and R.~Gatto,
  Phys.\ Lett.\  B {\bf 155} (1985) 95;

\bibitem{Casalbuoni:1986vq}
 R.~Casalbuoni, S.~De Curtis, D.~Dominici and R.~Gatto,
  Nucl.\ Phys.\  B {\bf 282} (1987) 235.

\bibitem{Georgi:1989xy}
  H.~Georgi,
  Nucl.\ Phys.\  B {\bf 331} (1990) 311.


  \bibitem{Giudice:2007fh} 
  G.~F.~Giudice, C.~Grojean, A.~Pomarol and R.~Rattazzi,
  JHEP {\bf 0706}, 045 (2007)
  [hep-ph/0703164].
 

\bibitem{Calibbi:2012at} 
  L.~Calibbi, Z.~Lalak, S.~Pokorski and R.~Ziegler,
  JHEP {\bf 1207}, 004 (2012)
  [\hhref{1204.1275}[hep-ph]].
  

\bibitem{Domenech:2012ai} 
  O.~Domenech, A.~Pomarol and J.~Serra,
  Phys.\ Rev.\ D {\bf 85}, 074030 (2012)
  [\hhref{1201.6510v2}[hep-ph]].


\bibitem{Gavela:2014vra}
  M.~B.~Gavela, J.~Gonzalez-Fraile, M.~C.~Gonzalez-Garcia, L.~Merlo, S.~Rigolin and J.~Yepes, JHEP {\bf 1410} (2014) 44
  [arXiv:1406.6367 [hep-ph]].


\bibitem{Yepes:2015qwa}
  K.~M.~Ruan, J.~Shu and J.~Yepes,
  Commun.\ Theor.\ Phys.\  {\bf 66} (2016) no.1,  93
  [arXiv:1507.04745 [hep-ph]].


\bibitem{Yepes:2015zoa}
  J.~Yepes,
  PoS CORFU {\bf 2015} (2016) 063
  [arXiv:1507.03974 [hep-ph]].
  
  
\bibitem{Alonso:2012pz}
  R.~Alonso, M.~B.~Gavela, L.~Merlo, S.~Rigolin and J.~Yepes,
  Phys.\ Rev.\ D {\bf 87} (2013) no.5,  055019
  [arXiv:1212.3307 [hep-ph]].


\bibitem{Alonso:2012px}
  R.~Alonso, M.~B.~Gavela, L.~Merlo, S.~Rigolin and J.~Yepes,
  Phys.\ Lett.\ B {\bf 722} (2013) 330
   Erratum: [Phys.\ Lett.\ B {\bf 726} (2013) 926]
  [arXiv:1212.3305 [hep-ph]].
  

\bibitem{Baker:2006ts}
C.~Baker, D.~Doyle, P.~Geltenbort, K.~Green, M.~van~der Grinten, {\em et.~al.},
  {\it {An Improved Experimental Limit on the Electric Dipole Moment of the
  Neutron}},  Phys.Rev.Lett. {\bf 97} (2006) 131801,
  [\href{http://xxx.lanl.gov/abs/hep-ex/0602020}{{\tt hep-ex/0602020}}].


  \bibitem{Panico:2011pw} 
  G.~Panico and A.~Wulzer,
  JHEP {\bf 1109}, 135 (2011)
  [\hhref{1106.2719}[hep-ph]].
  
  \bibitem{DeCurtis:2011yx} 
  S.~De Curtis, M.~Redi and A.~Tesi,
  JHEP {\bf 1204}, 042 (2012)
  [\hhref{1110.1613}[hep-ph]].


\bibitem{Falkowski:2011ua}
  A.~Falkowski, C.~Grojean, A.~Kaminska, S.~Pokorski and A.~Weiler,
  JHEP {\bf 1111} (2011) 028
  [\hhref{1108.1183} [hep-ph]].
 

\bibitem{Contino:2006nn}
  R.~Contino, T.~Kramer, M.~Son, R.~Sundrum,
  JHEP {\bf 0705} (2007) 074
  [\hhref{hep-ph/0612180} [hep-ph]].

\bibitem{Redi:2011zi}
  M.~Redi, A.~Weiler,
  JHEP {\bf 1111} (2011) 108
   [\hhref{1106.6357} [hep-ph]].


\bibitem{Chatrchyan:2012rva}
  S.~Chatrchyan {\it et al.} [CMS Collaboration],
  JHEP {\bf 1302} (2013) 036,  [arXiv:1211.5779 [hep-ex]].
  

\bibitem{CMS:2013qca}
  CMS Collaboration,
  CMS-PAS-EXO-12-061.

\bibitem{ATLAS:2013jma}
 ATLAS Collaboration,
  ATLAS-CONF-2013-017.


\bibitem{CMS:2013vda}
  CMS Collaboration,
  CMS-PAS-EXO-12-025.

\bibitem{ATLAS:2013lma}
ATLAS Collaboration,
  ATLAS-CONF-2013-015.
  

\bibitem{CMS:2013xea}
  CMS Collaboration,
  CMS-PAS-EXO-12-021.

\bibitem{CMS:2013dfa}
  CMS Collaboration,
  CMS-PAS-EXO-12-022.
  
\bibitem{CMS:2013fea}
  CMS Collaboration,
  CMS-PAS-EXO-12-024.
  

\bibitem{Pappadopulo:2014qza}
  D.~Pappadopulo, A.~Thamm, R.~Torre and A.~Wulzer,
  JHEP {\bf 1409} (2014) 060
  [arXiv:1402.4431 [hep-ph]].
  
\bibitem{Greco:2014aza}
  D.~Greco and D.~Liu,
  JHEP {\bf 1412} (2014) 126
  [arXiv:1410.2883 [hep-ph]].


\bibitem{Shu:2016exh}
  J.~Shu and J.~Yepes,
  Mod.\ Phys.\ Lett.\ A {\bf 31} (2016) no.40,
  [arXiv:1601.06891 [hep-ph]].


\bibitem{Shu:2015cxm}
  J.~Shu and J.~Yepes,
  Commun.\ Theor.\ Phys.\  {\bf 66} (2016) no.6,  643
  [arXiv:1512.09310 [hep-ph]].


  \bibitem{Chatrchyan:2013wfa}
  S.~Chatrchyan {\it et al.} [CMS Collaboration],
  Phys.\ Rev.\ Lett.\  {\bf 112} (2014) no.17,  171801
  [arXiv:1312.2391 [hep-ex]].
  
  
\bibitem{Chatrchyan:2013uxa}
  S.~Chatrchyan {\it et al.} [CMS Collaboration],
  Phys.\ Lett.\ B {\bf 729} (2014) 149
  [arXiv:1311.7667 [hep-ex]].
  

\bibitem{ATLAS:2014pga}
ATLAS collaboration,
  ATLAS-CONF-2014-036.
  

\bibitem{Azatov:2013hya} 
  A.~Azatov, M.~Salvarezza, M.~Son and M.~Spannowsky,
  Phys.\ Rev.\ D {\bf 89}, 075001 (2014)
  [\hhref{1308.6601}[hep-ph]].
  
\bibitem{Matsedonskyi:2014mna} 
  O.~Matsedonskyi, G.~Panico and A.~Wulzer,
[\hhref{1409.0100}[hep-ph]].
  

\bibitem{Matsedonskyi:2015dns}
  O.~Matsedonskyi, G.~Panico and A.~Wulzer,
  JHEP {\bf 1604} (2016) 003
  [arXiv:1512.04356 [hep-ph]].


\bibitem{Matsedonskyi:2014lla} 
  O.~Matsedonskyi, F.~Riva and T.~Vantalon,
  JHEP {\bf 1404}, 059 (2014)
  [\hhref{1401.3740}[hep-ph]].
      
\bibitem{Aaboud:2017qpr}
  M.~Aaboud {\it et al.} [ATLAS Collaboration],
  JHEP {\bf 1708} (2017) 052
  [arXiv:1705.10751 [hep-ex]].      
      
      
 \bibitem{Sirunyan:2017pks}
  A.~M.~Sirunyan {\it et al.} [CMS Collaboration],
  Phys.\ Lett.\ B {\bf 779} (2018) 82
  [arXiv:1710.01539 [hep-ex]].   
    
\end{thebibliography}
\end{document}